\documentclass[runningheads,a4paper,oribibl]{llncs}

\usepackage{rotating}
\usepackage{graphicx}

\usepackage{amssymb}
\usepackage{graphicx}
\usepackage[table]{xcolor}
\usepackage[font=small,labelfont=bf]{caption}
\usepackage{graphicx}
\usepackage{caption}
\usepackage{amsmath}

\usepackage{url}
\usepackage{float}

\newcommand{\ignore}[1]{}

\usepackage{url}

\urldef{\mailsa}\path|imrul@mail.usf.edu,jacob@ua.edu|

\begin{document}




%
%
%


\title{The Network of Faults: A Complex Network Approach to Prioritize Test Cases for Regression Testing}
\titlerunning{Prioritizing Test Cases for Regression Testing}

\author{Imrul Kayes\inst{1}
\and Jacob Chakareski \inst{2}\\
}
\institute{Computer Science \& Engineering \\University of South Florida, Tampa, FL, USA \and
Electrical and Computer Engineering\\
University of Alabama, Tuscaloosa, Alabama, USA
\mailsa}

\maketitle

%
%


\begin{abstract}
Regression testing is performed to provide confidence that changes in a part of software do not affect other parts of the software. 
An execution of all existing test cases is the best way to re-establish this confidence.
However,  regression testing is an expensive process---there might be insufficient resources (e.g., time, workforce) to allow for the re-execution of all test cases.
Regression test prioritization techniques attempt to re-order a regression test suite based on some criteria so that  highest priority test cases are executed earlier.

In this study, we want to prioritize test cases for regression testing based on the dependency network of faults.
In software testing, it is common that some faults  are consequences of other faults (leading faults).
Moreover, dependent faults can be removed if and only if the leading faults have been removed.
Our goal is to prioritize test cases so that test cases that exposed leading faults (the most central faults in the fault dependency network) in the system testing phase, are executed first in regression testing.

We present ComReg, a test case prioritization technique based on the dependency network of faults.
We model a fault dependency network as a directed graph and identify leading faults to prioritize test cases for regression testing.
We use a centrality aggregation technique which considers six network representative centrality metrics to identify leading faults in the fault dependency network.
We also discuss the use of fault communities to select an arbitrary percentage of the test cases from a prioritized regression test suite.
We conduct  a case study that evaluates the effectiveness and applicability of the proposed method.

We obtain a  fault dependency network from the development of a vocabulary learning software.
We found that the fault network is a small-world graph with distinguishable community structure.
The leading faults are common in all centralities and a re-ordering of test cases is feasible for regression testing based on those leading faults.
Our method outperforms traditional regression testing prioritization techniques in detecting fault dependencies.
Our  modeling of the network of faults provides insights into the requirement of recognizing fault dependencies while re-ordering regression test suites  for both research and practice. 
The  dependency model needs further evaluation and improvement considering associated resources (e.g., man-hours).


\end{abstract}

\section{Introduction}

Regression testing is performed after a software is modified. 
The purpose of regression testing is to test the modified software with some test cases in order to re-establish our confidence that the software will perform according to the modified specification and the newly introduced changes do not hinder the behavior of the unchanged part of the software.
In a development cycle, regression testing may begin after the detection and correction of faults in a tested software~\cite{LeungRegresion}.
Regression test suite ensures that the evolution of an application does not result in a low quality software product.
However, regression testing has become a complex procedure  because of  recent trends in software development paradigms. 
For example, short and iterative ``Agile'' software development  imposes restrictions and constraints on how regression testing can be performed within limited resources~\cite{yoo2012regression}.

Intuitively, the best way to re-gain confidence from regression testing is to  execute all existing test cases from a test suite.
Unfortunately, regression testing is often directly associated with high costs.
Beizer~\cite{beizer2003software} points out that regression testing accounts for as much as one-half the cost of software maintenance.
One industrial collaborator of Elbaum et al.~\cite{malishevsky2006cost} reports that for one of their products of about 20,000 lines of code, the entire test suite requires seven weeks to run.
Some of the most well-studied software failures, for example, the Ariane-5 rocket was blamed on the failure to test changes in a software system~\cite{Hamlet2000Software}.

In general, test case prioritization techniques seek to schedule test cases in an order so that the tester obtains maximum benefit, even if the testing is prematurely halted at some arbitrary point~\cite{yoo2012regression}. 
Regression test prioritization aims to re-order a regression test suite so that those tests with highest priorities, according to some established criterion, are executed earlier in the process of regression testing  than those with lower priorities~\cite{Elbaum2002Pri}.
Researchers have proposed various techniques for test case prioritization to re-order the test cases for regression testing.
These techniques focus on various aspects of product development, such as coverage-based approaches~\cite{Rothermel1999Prio,Rothermel2001Pri,Elbaum2000PTC,Elbaum2002Pri}, requirement-based approaches~\cite{Srikanth2005Requirement,krishnamoorthi2009factor} and constraint-based approaches~\cite{Zhang2009TTP,Walcott2006TTS,Alspaugh2007ETP}.

However, none of the solutions addressed dependencies among faults while prioritization.
In software testing, it is known that some faults  are the consequences of other faults (commonly termed as leading faults).
Experience shows that in a software development process, mutually independent faults can be directly detected and removed, but dependent faults can be removed if and only if leading faults have been removed~\cite{Huang2006Fault}.
In worst cases, fault dependencies can create a cascade of faults that can severely effect a software system. 
For example, in 1990, a fault in the failure recovery code of the AT\&T led to cascading faults, which costs 9 hours of downtime and at least $60$ million in lost revenue~\cite{neumann1990cause}. 
Another example of cascading faults is the escalation of a divide-by-zero exception into a Navy ship's network that left the smart ship dead in the water~\cite{slabodkin1998software}. Researchers hint that the Internet is also at risk of cascading failures~\cite{oppenheimer2003internet}.
We argue that test case that reveals leadings faults should be executed first in a regression testing process in order to get an early confirmation that the software is free from dependent faults.

We attempted a first step to prioritize regression testing based on fault dependency in~\cite{Kayes2011Fault}.  
We proposed an algorithm to prioritize test cases based on fault dependency. 
However, in~\cite{Kayes2011Fault}, we only considered $1$-hop neighborhood or dependencies of faults.
This paper uses a fault dependency network to prioritize test cases for regression testing.
We leverage faults' positions in the network to determine leading faults (central faults in the network).

The contributions of this work are:
\begin{itemize}
  \item First, we describe ComReg, which leverages fault dependency network to prioritize test cases for regression testing.
   We present a directed graph model for the fault dependency network and identify leading faults (central faults) to prioritize test cases.
   Our identification of  leading faults  is based on a centrality aggregation technique.
   Centralities can represent the position of a fault in a fault network.
   We propose an aggregation of different representative centrality metrics (indegree, betweenness, closeness, eigenvector, pagerank, and hub centrality) into a final leading score to identify leading faults.
   
 \item Second, we discuss the use of fault communities to select $X\%$ of the test cases from a prioritized regression test suite.
  
  \item Finally, we present a case study from the  development of a subject software ``Tarantula''.
 We discuss the test cases written for the software, the faults it exposed after testing, and the fault network from the exposed faults.
 We show the identification of leading faults  for prioritization and compare the effectiveness with traditional techniques.
 We also show fault communities for a selection of  test cases from the prioritized regression test suite.

  \end{itemize}
  
The rest of the paper is organized as follows. 
Section~\ref{method} introduces fault dependency-aware test case prioritization technique, ComReg. 
Section~\ref{casestudy} presents a case study. 
Section~\ref{related} reviews related work and Section~\ref{conclusion} concludes. 

\section{Fault Dependency-Aware Test Case Prioritization}
\label{method}

\subsection{Problem Statement}
Based on Elbaum et al.~\cite{Elbaum2000PTC}, we define a prioritization of test cases for regression testing as follows.

Given  a test suite $T$, the set of permutations of $T$ as $PT$ and a function from $PT$ to the real numbers as $f$,
a prioritization of test cases for regression testing solution provides an ordered test suite $T'$ such that for all $T''$, $f(T') \geq f(T'')$, where $T' \in PT$ and $T'' \in PT$.

$PT$ represents the set of all possible prioritization (orderings) of $T$ and $f $ is an utility function that, applied to any such ordering, yields an award value to that ordering. 
For example, let us we have $n$ test cases as $(T_1, T_2, T_3, \dots ,T_n) \in T$. 
From those test cases, $n!$ orderings are possible.
Test case prioritization techniques attempt to find an order from $n!$ number of orderings such that the order maximizes the utility function $f$.

Let us we have $t$ test cases $(T_1, T_2, T_3, \dots ,T_t)$ in the test suite $T$.
After running those test cases for a system testing, we get $n$ faults such as $(F_1, F_2, F_3, \dots ,F_n)$ as $F$.
There exists a relation from $T$ to $F$, $R:T \rightarrow F$, such that for each test case $t \in T$ there exists none, single or multiple faults $f \in F$.
Our goal is to prioritize the test suite and select $X\%$ of the test cases for regression testing.

\subsection{Our Approach-ComReg}
\label{comreg}

We propose ComReg, a fault dependency-aware test case prioritization for regression testing.
ComReg is based on the fact that mutually independent faults can be directly detected and removed, but dependent faults can be removed if and only if the leading faults have been removed~\cite{Huang2006Fault}.
A leading fault is the fault that causes dependent faults to occur.
For example, consider a simple dictionary program, which has a \textit{load} functionality to read all words and their meanings from text files, a \textit{next word} functionality that allows users to browse words and a \textit{random number generator} for generating a number for an arbitrary selection of a word list.
The  \textit{next word} functionality is dependent on  the \textit{load} functionality in that if the system fails to read words and meanings, there is no way to browse the words.
So, consider three following faults that occur.

\begin{enumerate}
  \item \textit{Fault F1}: words and meanings upload failure.
  \item \textit{Fault F2}: does not find the next word.
   \item \textit{Fault F3}: random generator does not show a random number.
\end{enumerate}

\begin{figure}[htbp]
\centering
\includegraphics[height=4.00cm]{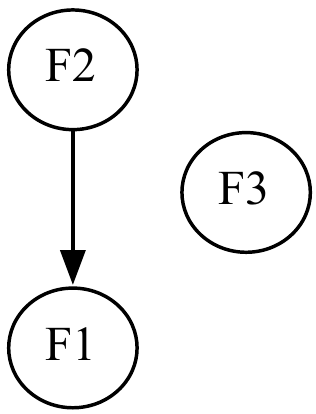}
\caption{Fault dependency.}
\label{fig:dependency1}
\end{figure}

Figure~\ref{fig:dependency1} shows the faults.
We can draw an arrow from fault F2 to fault F1 to show the dependency of F2 on F1.
In this case, F1 is a leading fault and F2 is a dependent fault.
However, the fault F3 is an independent fault (no arrow to or from F3 in  Figure~\ref{fig:dependency1}).
Leading faults might be limited in numbers.
For example, Microsoft  reports that 80 percent of the errors and crashes in Windows and Office are caused by 20 percent of the entire pool of faults~\cite{microsoftfaults}.
We propose to prioritize a regression test suite based on leading faults and to run $X\%$ of the test cases which contain leading faults.
So, in our fault dependency-aware test case prioritization, an order of the test cases attempts to maximize the utility function $f$ that determines the number of the leading faults.
Fault dependencies could be  data, control or module dependent.

\begin{figure*}[htbp]
\centering
\includegraphics[height=6.00cm]{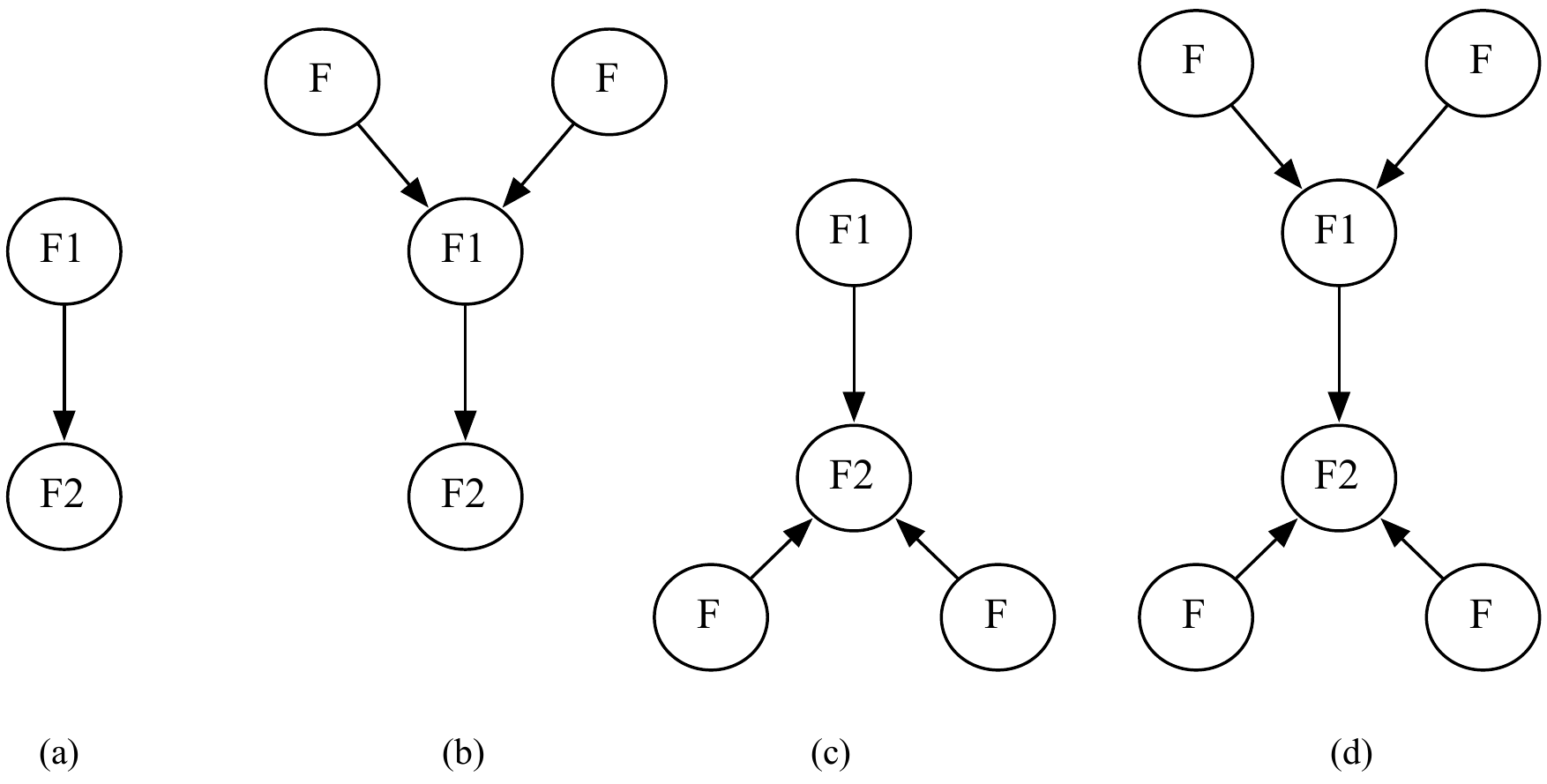}
\caption{Examples of various fault dependencies considering two faults: Fault F1 and Fault F2.  (a) Fault F1 is dependent on Fault F2 (b) Fault F1 is dependent on Fault F2 and other faults are dependent on Fault F1 (c) Fault F1 is dependent on Fault F2 and other faults are dependent on Fault F2 (d) Fault F1 is dependent on Fault F2 and other faults are dependent on both Faults F1 and F2.}
\label{fig:sample_dependency}
\end{figure*}

However, the problem of detecting leading faults is not trivially solvable.
The challenge is due to the fact that faults not only have local effects (e.g., Fault A is dependent on fault B, so A could not be removed before removing B), but also faults have global effects too (e.g., Fault A is dependent on fault B and Fault B is dependent on Fault C. Fault A could not be removed before removing Fault B and Fault C). 
We present various scenarios of fault dependencies considering two faults F1 and F2 in Figure~\ref{fig:sample_dependency}.
Some of the scenarios are listed below.

\begin{itemize}
\item Fault F1 is dependent on Fault F2, or vice versa
\item a) Fault F1 is dependent on Fault F2; b) other faults are dependent on Fault F1; c) vice versa of (a) and (b)
\item a) Fault F1 is dependent on Fault F2; b) other faults are dependent on Fault F2; c) vice versa of (a) and (b)
\item a) Fault F1 is dependent on Fault F2; b) other faults are dependent on Fault F1; c) other faults are dependent on Fault F2 (d) vice versa of (a), (b) and (c)
\end{itemize}

\begin{figure*}[htbp]
\centering
\includegraphics[height=10.00cm]{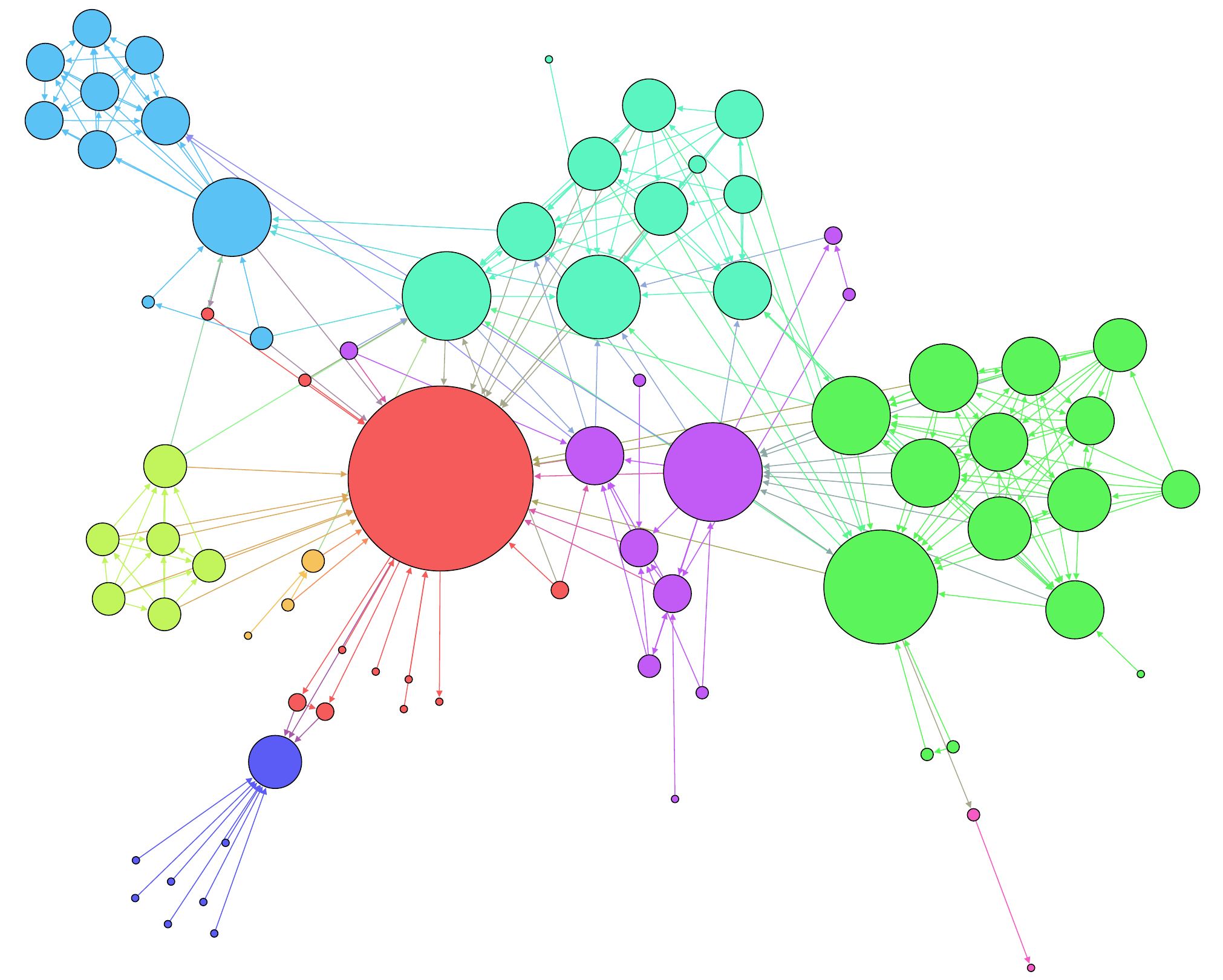}
\caption{A  fault dependency network  of $77$ nodes and $254$ dependencies. Node size is proportional to in-degree. }
\label{fig:a_dependencyGraph}
\end{figure*}

So, it appears that if we consider all faults and their dependencies, the situation becomes very complex.
All faults and their dependencies can be captured by a complex network as shown in Figure~\ref{fig:a_dependencyGraph}.
The network is comprised of $77$ faults (shown as nodes) and $254$ dependencies (shown as edges).
There is an edge from Fault A to Fault B if Fault A is dependent on Fault B.
The leading faults in the network are those who occupy central positions.

Formally, we model a fault dependency network as a directed graph $F=(V,E)$, where a node $v \in V$ is a fault and an edge $e_{ij} \in E$ from $v_i \in V$ to $v_j \in V$ denotes that the fault $v_i$ is dependent on the fault $v_j$.
The number of nodes and edges are $|V|=n$ and $|E|=m$ respectively.
The directed graph can be represented by a $n * n$ matrix $F_{n*n}$, where an entry $F(i,j):$

\begin{equation}
F(i,j)=\begin{cases}
   1  &\text{if $e_{ij} \in E$}\\
    0, & \text{otherwise}
  \end{cases}
\end{equation}

The position of a node (fault) in the network can be represented in network analysis by different centrality metrics.
For example, the larger the number connections a fault receive from its direct neighbors, the higher number of other faults depend on the fault.
Without removing the faults all dependent faults could not be removed.
Alternatively, the larger the number of paths between other pairs of faults a fault is part of, the more it can control the fault propagation between distant faults.
Based on this intuition, we conjecture that a fault's position is determined by and manifests via its centrality in the fault network. 

We propose to aggregate different representative centrality metrics into a final leading score to identify the leading faults based on ~\cite{Imrul12Influential}. 
In~\cite{Imrul12Influential}, the authors used centrality aggregation technique to identify influential bloggers in a blogging network.
We define the leading score of a fault (node) as the average of the positions of that node in decreasing order of centrality scores over various centrality metrics. 
Specifically, each centrality metric assigns each node a score that can be used to order the nodes in decreasing order of importance (according to that centrality). 
This allows each fault to receive a rank according to each centrality metric: the first ranked fault will be the most central one, the last ranked will be the one with the lowest centrality score. 
Faults having the same centrality score are given the same rank. 
A fault's final rank is the average rank over all centrality measures. 
We selected six representative centrality metrics as the focus of our study: indegree, betweenness, closeness, eigenvector, pagerank, and hub centrality. 


Degree centrality is defined as the number of links that a node has. 
In a directed graph like fault dependency graph, two types of degree centralities are possible: indegree and outdegree centrality.
For a node, the number of  direct incoming connections is characterized as indegree of the node.
On the other hand, the number of  direct outgoing connections is characterized as out degree of the node.
Although simple, indegree centrality intuitively captures an important aspect of a fault's potential leading position: faults who have many incoming connections from many other faults are those that make other faults to depend. 
In our fault dependency graph $F$, the indegree centrality of a fault $i$ can be represented by the following equation.
\begin{equation}
indegree(i)=\sum_{1 \leq j \leq n} F_{ji}
\end{equation}

Betweenness centrality, which measures the extent to which a node lies on the shortest paths between other nodes, was introduced as a measure for quantifying the control of a human on the communication between other humans in a social network~\cite{Freeman1977Betweenness}.
Faults with high betweenness centrality may have considerable influence within a fault dependency network by virtue of their control over fault propagation among other faults.
The nodes with the highest betweenness are also the ones whose removal from the network will most disrupt communications between other nodes because they lie on the largest number of paths taken by faults~\cite{Newman2010Book}. 
Formally, the betweenness centrality of a node is the sum of the fraction of all-pairs shortest paths that pass through :
\begin{equation}
C(v)=\sum_{s,t\in V}\frac{\sigma(s,t|v)}{\sigma(s,t)}
\end{equation}
where v is the set of nodes, $\sigma (s, t)$ is the number of shortest $(s, t)$ paths, and $(s, t| v)$ is the number of those paths passing through some nodes $v$ other than $s, t.$ If $s=t$, $\sigma(s, t)=1,$ and if $v\in s, t, \sigma(s, t| v)=0$. 
Our implementation of betweenness for this research is based on the Brandes algorithm~\cite{Brandes2001BetweennessVariantAlgo}. 

Closeness centrality measures the mean distance from a node to other nodes, assuming that faults propagate along the shortest paths. 
Formally, the closeness centrality $(C(x))$ of a node $x$ is defined as follows:
\begin{equation}
C(x)= \frac{n-1}{\sum_{y\in U, y \neq x }d(x,y)}     \end{equation}
where $d(x, y)$ is the distance between node $x$ and node $y$; $U$ is the set of all nodes; $d$ is the average distance between $x$ and the other nodes. 
In our fault dependency network, this centrality measure estimates the amount of faults a fault may have access to compared to other faults. 
Specifically, a fault with lower mean distance to others can reach others faster. 


The centrality of a node does not only depend on the number of its adjacent nodes, but also on their relative importance. 
Eigenvector centrality allocates relative scores to all nodes in the network such that high-score neighbors contribute more to the score of the node.
Formally, Bonacich~\cite{BonacichClique} defines the eigenvector centrality $C(v)$ of a node $v$ as the function of the sum of the eigenvector centralities of the adjacent nodes, i.e.
\begin{equation}
C(v)=1/\lambda \sum_{(v,t)\in E}c(t)
\end{equation}
where $\lambda$ is a constant. This can be rewritten in vector notation, resulting in an eigenvector equation with well-known solutions. 

Originally designed as an algorithm to rank web pages~\cite{Page1999Pagerank}, PageRank computes a ranking of the nodes in a graph based on the structure of the incoming links.
The algorithm  assigns a numerical weighting to each node of a network  with the purpose of ``measuring'' its relative importance within the network.

Hubs and authorities are other relevant centralities for the fault network context. 
In a graph, authorities are nodes that contain useful information on a topic of interest; hubs are nodes that know where the best authorities are to be found~\cite{Newman2010Book}. 
A high authority centrality node is pointed to by many hubs, i.e., by many other nodes with high hub centrality. 
A high hub centrality node points to many nodes with high authority centrality. 
These two centralities can play a significant role also in our work of finding leading faults. 
They can infer that the faults that have high hub and authority centrality are not only leading but also they are connected with leading faults.

\subsection{Fault Communities to Select $X\%$ of Test Cases }

A common approach of regression testing is to select and run $X\%$ of the test cases from a prioritized test suite.
However, an optimal selection is always challenging.
On one hand, a few selections of test cases might remain a significant portion of the software virtually untested.
On the other hand, too many selections of test cases will require to test the entire system again.
However, in a fault network, fault communities could be leveraged to select an $X\%$ of the test cases.
Complex networks show communities in them: a community is a subset of nodes within which node to node connections are dense, but between which connections are less dense~\cite{girvan2002community}.
Communities are natural outcomes of real-world networks.
For example, e-mail network~\cite{tyler2005mail}, social application network~\cite{Nazir2008Apps}, mobile communication network~\cite{onnela2007structure}, blogging network~\cite{Kumar2004Blog}, and yeast protein-protein interaction network~\cite{chen2006detecting} revealed community structures.
Figure~\ref{fig:a_dependencyGraph} shows communities in a fault network; nodes in a community are colored the same.

Newman proposed a community detection algorithm~\cite{newman2004finding} based on modularity maximization.
Modularity is a utility function that computes the quality of a particular division of a network into communities. 
It is defined as the fraction of the edges that fall within the given community minus the expected such fraction if the edges were distributed at random.

\begin{equation}
Q = (E_1-E_2)
\end{equation}
where $E_1$= fraction of edges within communities and $E_2$=expected fraction of such edges.

The expected fraction of edges is typically evaluated within  a random graph conditioned on the degree sequence of the original network.
In that random graph, the probability of an edge between two
nodes $i$ and $j$ is $(k_i*k_j)/2m$, where $k_i$ is the degree of node $i$ and $m$ is the total number of edges in the network.
The modularity can then be written
\begin{equation}
Q = \frac{1}{2m}  \sum_{ij} \left ({A_{ij}- \frac{k_i*k_j}{2m}} \right) \delta(c_i,c_j)
\end{equation}

\begin{equation}
 \delta(c_i,c_j)=\begin{cases}
   1  &\text{if if i and j belong to the same community}\\
    0, & \text{otherwise}
  \end{cases}
\end{equation}

where $A_{ij}$ is the matrix representation of the graph, $\delta$ is the Kronecker delta, $c_i$ is the label of the community to which node $i$ is assigned.

The authors describe the modularity for an undirected graph.
However, the modularity can be extended for a directed graph such as fault dependency network.
In a random directed graph, the probability of an edge from node $j$ to node $i$ is $(k_{i}^{out}*k_{i}^{in})/m$.
Then for the fault dependency network the above equation could be written as

\begin{equation}
Q = \frac{1}{m}  \sum_{ij} \left ({F_{ij}- \frac{k_{i}^{out}*k_{i}^{in}}{m}} \right) \delta(c_i,c_j)
\end{equation}

Where $F$ is a fault dependency matrix and $F_{ij}$ is  1 if there is an edge from $j$ to $i$ and zero otherwise.

We propose to apply the community detection algorithm to uncover communities of the faults.
After detecting communities, the faults in  the same communities with a leading fault could be identified and corresponding test cases could be executed as a regression test.
Moreover, all modules in a software are not the same in terms of fault tolerance.
For example, login credential authentication or a module that processes financial transaction are more crucial than a module that prints documents.
Furthermore, Pareto principle also (known as 80-20 rule) applies to software systems.
The Pareto principle~\cite{pareto1927manual} states that for many events, roughly 80\% of the effects come from 20\% of the causes.
The Standish Group's report shows that in a software system, 45\% of features are never used, 19\% of features are rarely used, 19\% of features are used sometimes, 13\% of features are used often and only 7\% of features are always used~\cite{Duong2009Standish}.
So, in sum, only 20\% of software features are often and always used.
It becomes apparent that ensuring quality of those 20\% of software features is vital. 
Fault communities could be leveraged to ensure the quality of prioritized features (e.g.,  20\% of software features).
The leading faults and faults from their communities revealed by the test cases (which target prioritized features) could be used in selecting regression test cases.
This way a regression testing can ensure a high customer satisfaction.

\section{Case Study and Evaluation}
\label{casestudy}

The goal of the case study is to prioritize a test suite of size $N$ and identify $X\%$ of the test cases from the suite for regression testing.
To accomplish the goal, we developed a medium-scale English vocabulary learning software, ``Tarantula''.
In the process of development, we wrote test cases, executed the test cases, applied our method ComReg to prioritize the test cases and identified $X\%$ of the test cases for regression testing.
In this section, we first give a detailed overview of the functionalities of  ``Tarantula''.
Then we describe the test cases and relevant faults that we have got after applying the test cases in an initial  product.
We explain the features of the Tarantula, so that readers can understand the underlying goals of the test cases.
Finally, we present the leading faults using our centrality aggregation method and compare ComReg with traditional approaches.
We also show and discuss the relevance of using community detection techniques in the fault network.  

\subsection{An Overview of Tarantula}
\begin{figure}[htbp]
\centering
\includegraphics[height=5.50cm]{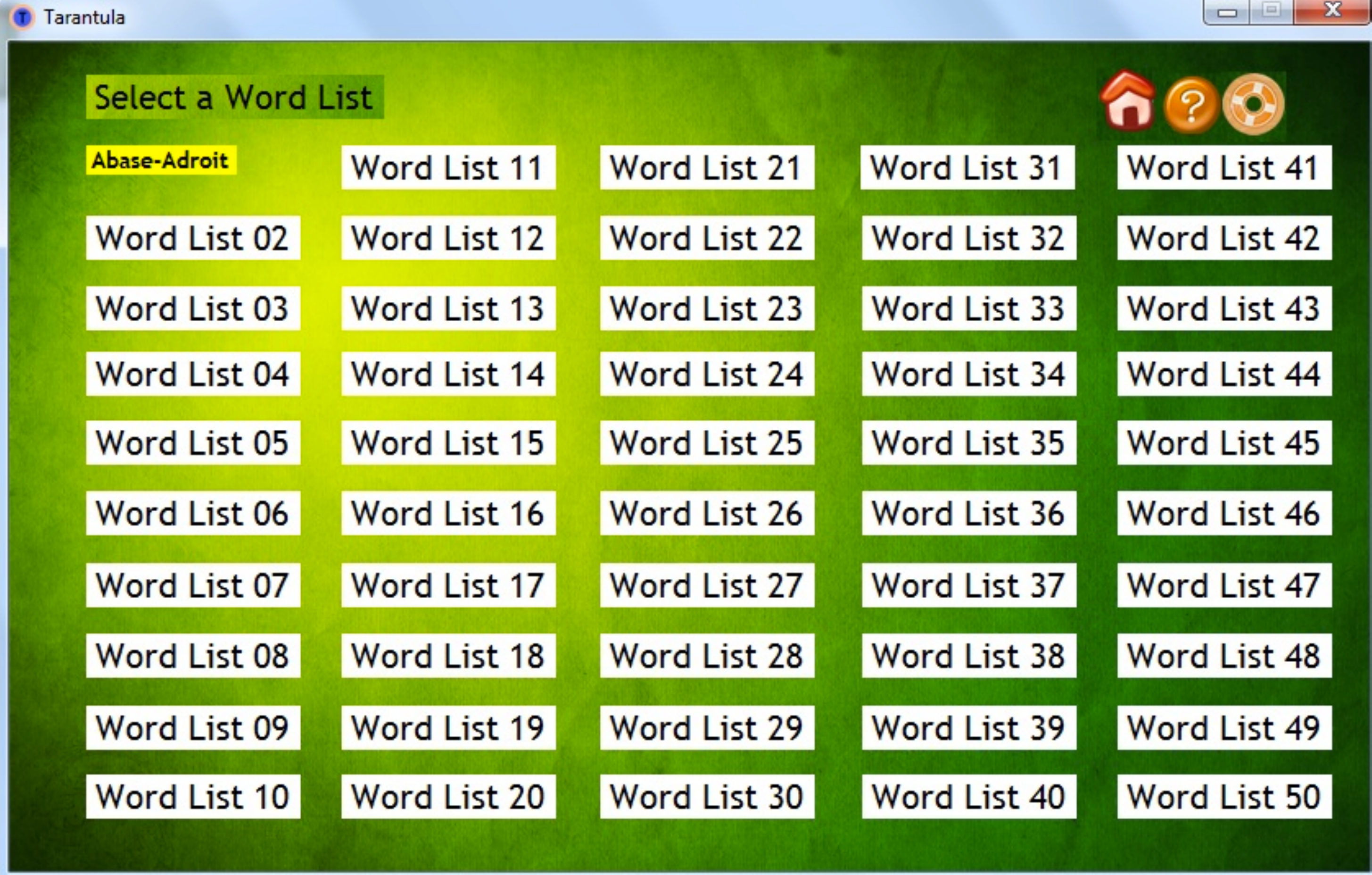}
\caption{First window of Tarantula.}
\label{fig:firstwindow}
\end{figure}

\begin{figure}[htbp]
\centering
\includegraphics[height=5.5cm]{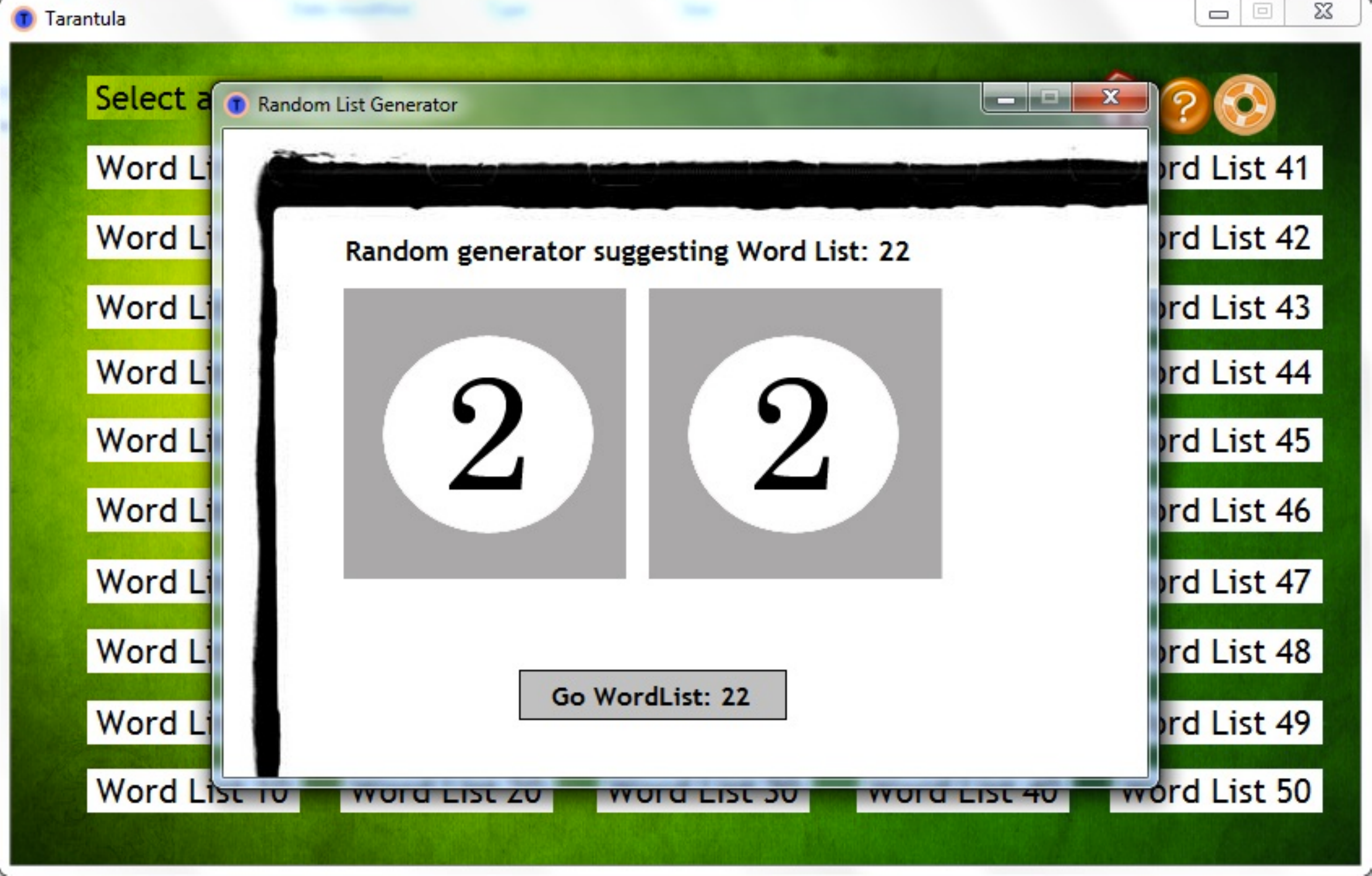}
\caption{Random window in Tarantula.}
\label{fig:random}
\end{figure}

\begin{figure}[htbp]
\centering
\includegraphics[height=5.50cm]{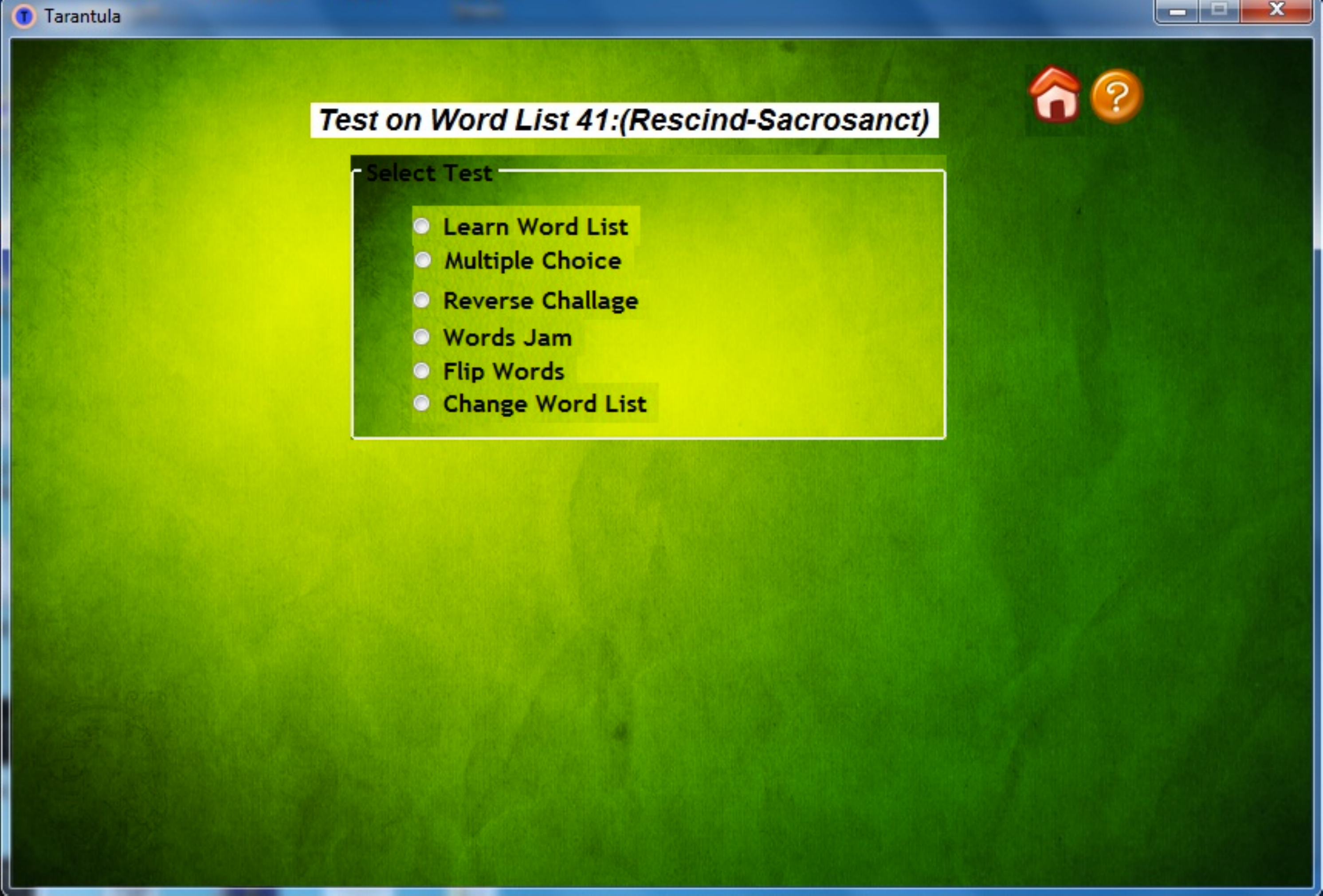}
\caption{Option window of Tarantula.}
\label{fig:optionwindow}
\end{figure}

\begin{figure}[htbp]
\centering
\includegraphics[height=5.50cm]{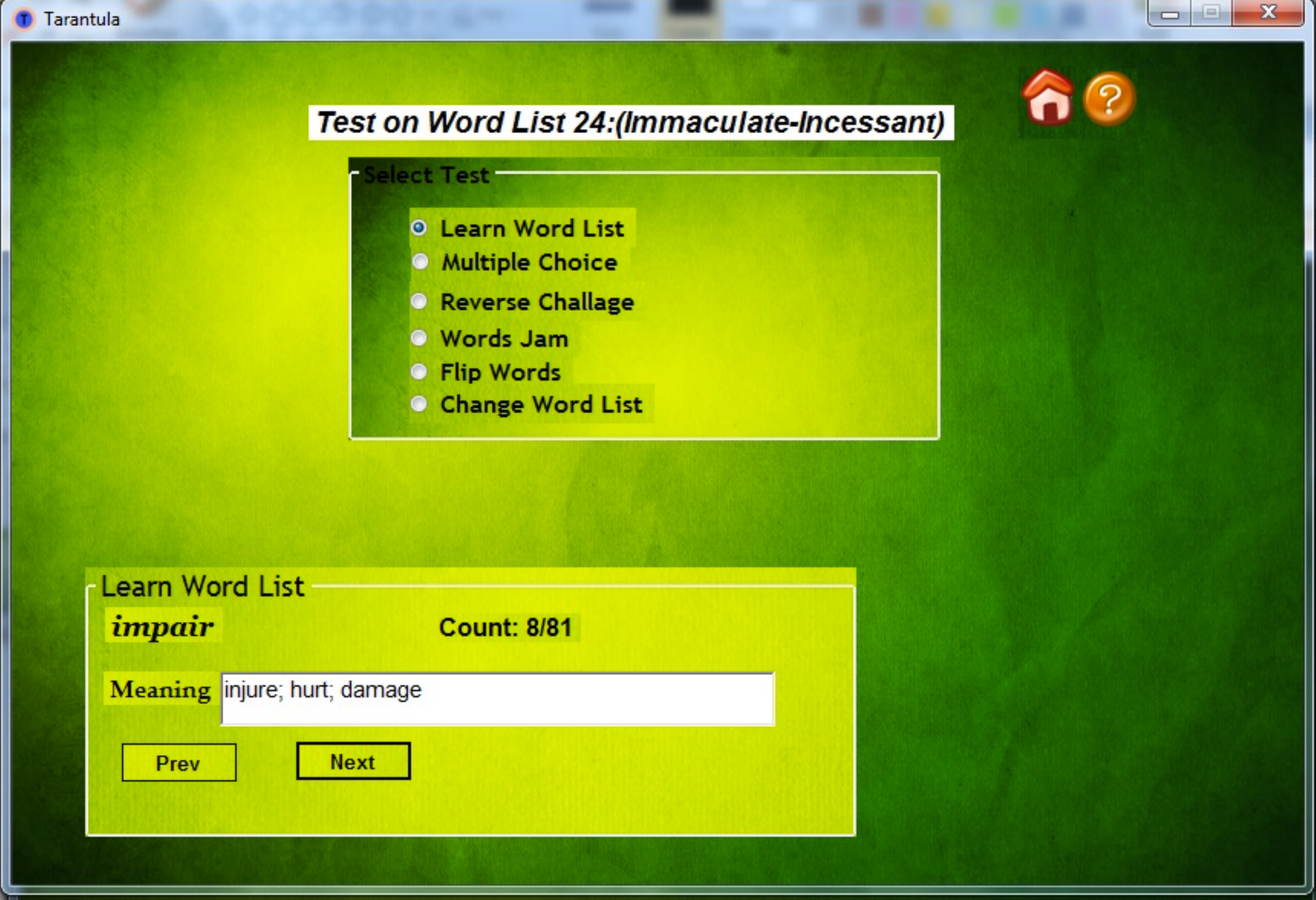}
\caption{Learn wordlist feature of Tarantula.}
\label{fig:learnwordlist}
\end{figure}

The Tarantula is an English vocabulary learning software, which is built targeting high frequency GRE words.
When a user installs and runs the software, she is presented with $50$ word lists.
Figure~\ref{fig:firstwindow} shows the first window of Tarantula.
The word lists consist of $4054$ words and when a user hovers a mouse on a word list icon, it shows the first and the last word in the word list.
Users can click on a word list icon to exercise various features of that word list.
However, users can use the random button (upper right corner in Figure~\ref{fig:firstwindow}) to select a random word list (see Figure~\ref{fig:random}).
When users select a word list, relevant features of the word list are shown in a separate option window as shown in Figure~\ref{fig:optionwindow}.
There are several options available on the option window for users to learn and practice words of a word list.
The options are: ``Learn WordList'', ``Multiple Choice'', ``Reverse Challenge'', ``Words Jam'' and ``Flip Words".
Users can click a radiobutton to select an option.
The ``Learn WordList'' feature shows the words and their meanings from the word list serially (see in Figure~\ref{fig:learnwordlist}).
Users can use ``Next'' and ``Prev'' buttons to view next and previous word respectively.
There is a counter which indicates the serial number or position of the word in the word list.

\begin{figure}[htbp]
\centering
\includegraphics[height=5.50cm]{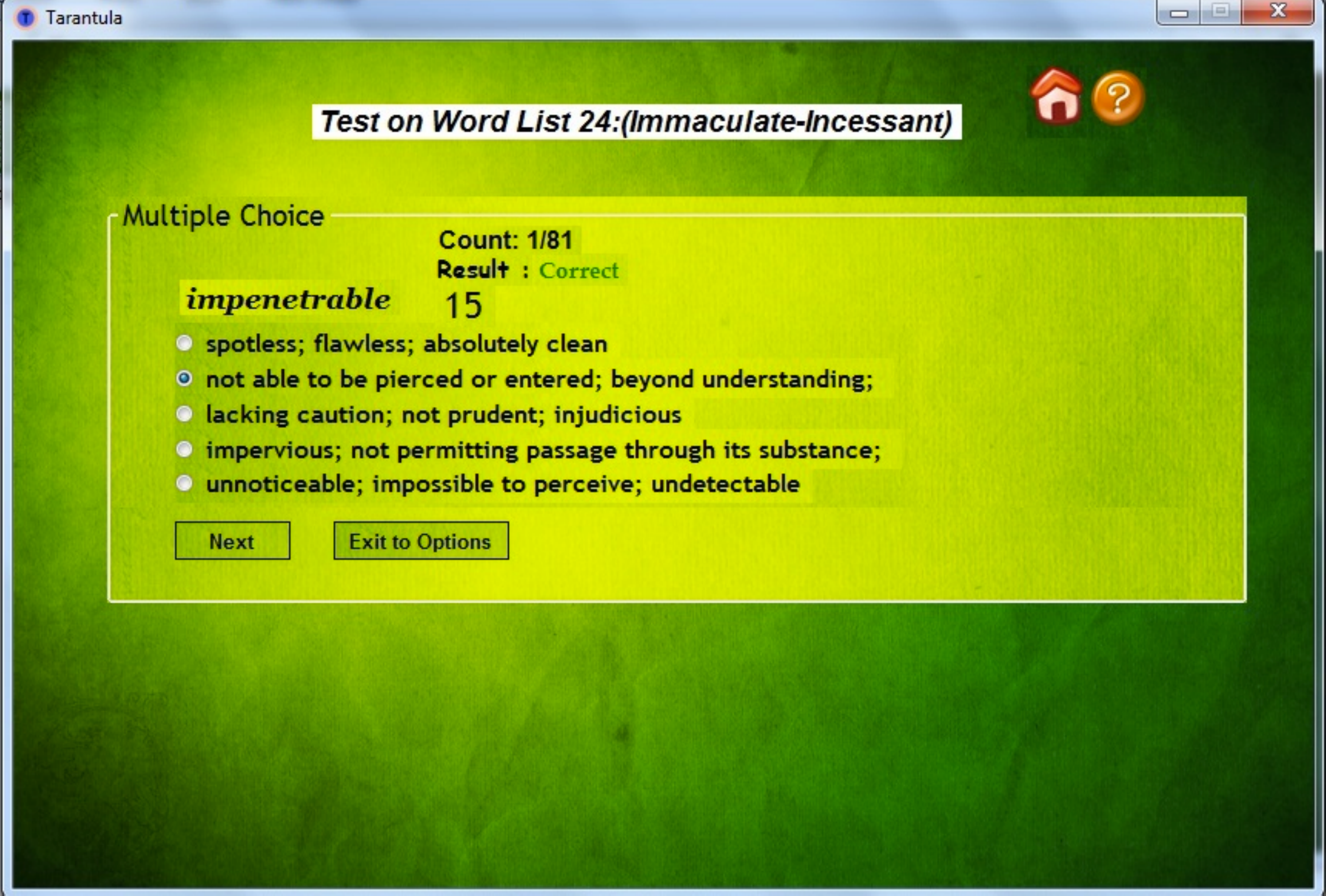}
\caption{Multiple Choice feature of Tarantula.}
\label{fig:multiple}
\end{figure}

\begin{figure}[htbp]
\centering
\includegraphics[height=5.50cm]{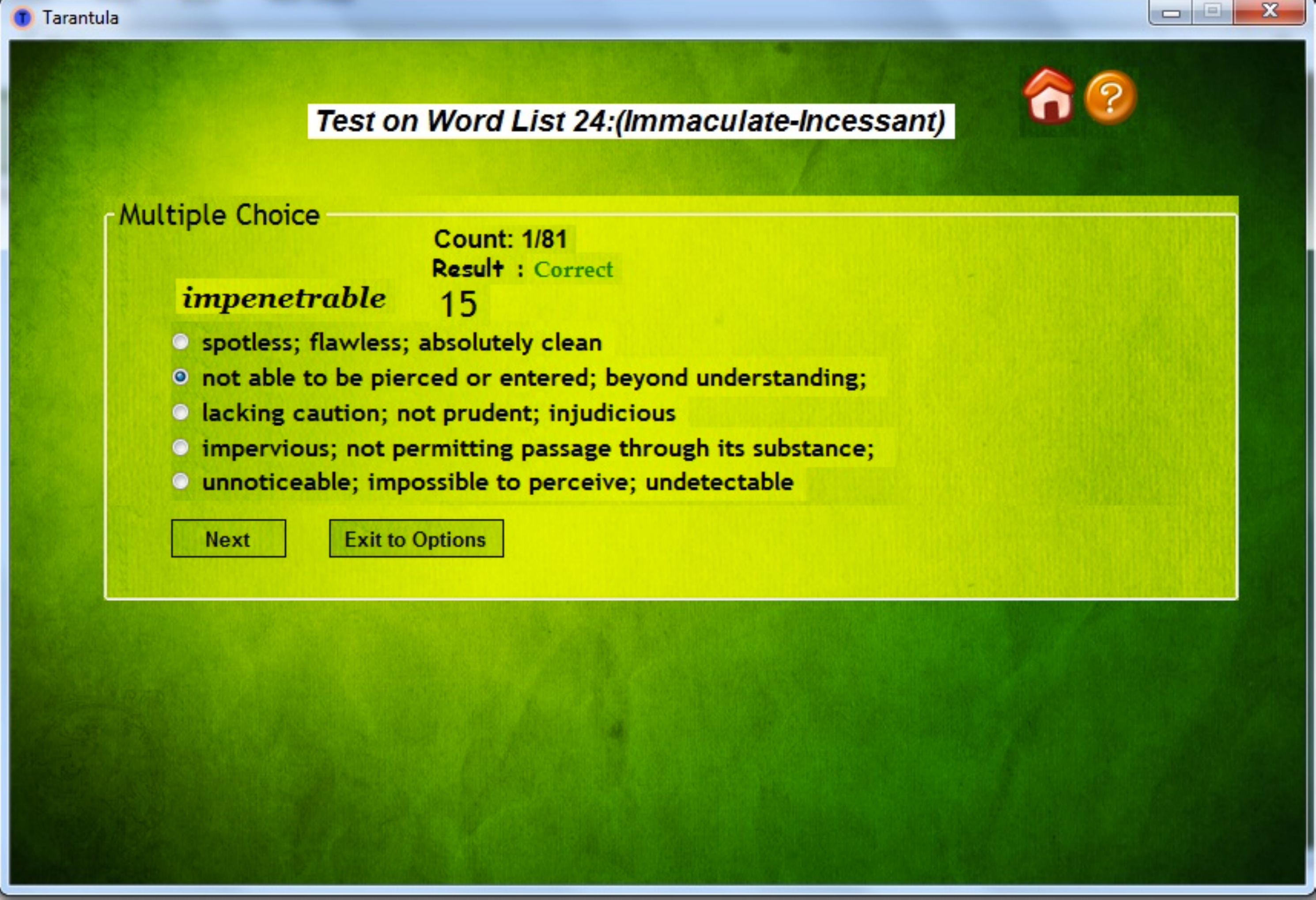}
\caption{Reverse Challenge Choice feature of Tarantula.}
\label{fig:reverse}
\end{figure}

The ``Multiple Choice'' feature shows a random word from the word list with five possible meanings (see Figure~\ref{fig:multiple}). 
Meanings are from the same word list taken randomly, and of them one is appropriate for the word shown.
When users click a  meaning of the word, the ``Result'' label shows whether the selection is right or wrong.
There is a timer label which increases in each second to show how much time a user is taking.
The ``Count'' label shows the number of words a user has practiced.
Users can click on a ``Next'' button to get a new word.
The ``Reverse Challenge'' feature is the opposite of ``Multiple Choice'' feature.
This  feature shows a random meaning from the word list with five possible words (see Figure~\ref{fig:reverse}). 
These words are from the same word list, taken randomly, and of them one is appropriate for the meaning shown.
When users click a  word of the meaning, the ``Result'' label shows whether the selection is right or wrong.
`Count'', `Next''  and ``Timer'' functionalities are similar to the functionalities of ``Multiple Choice'' feature.

\begin{figure}[htbp]
\centering
\includegraphics[height=5.50cm]{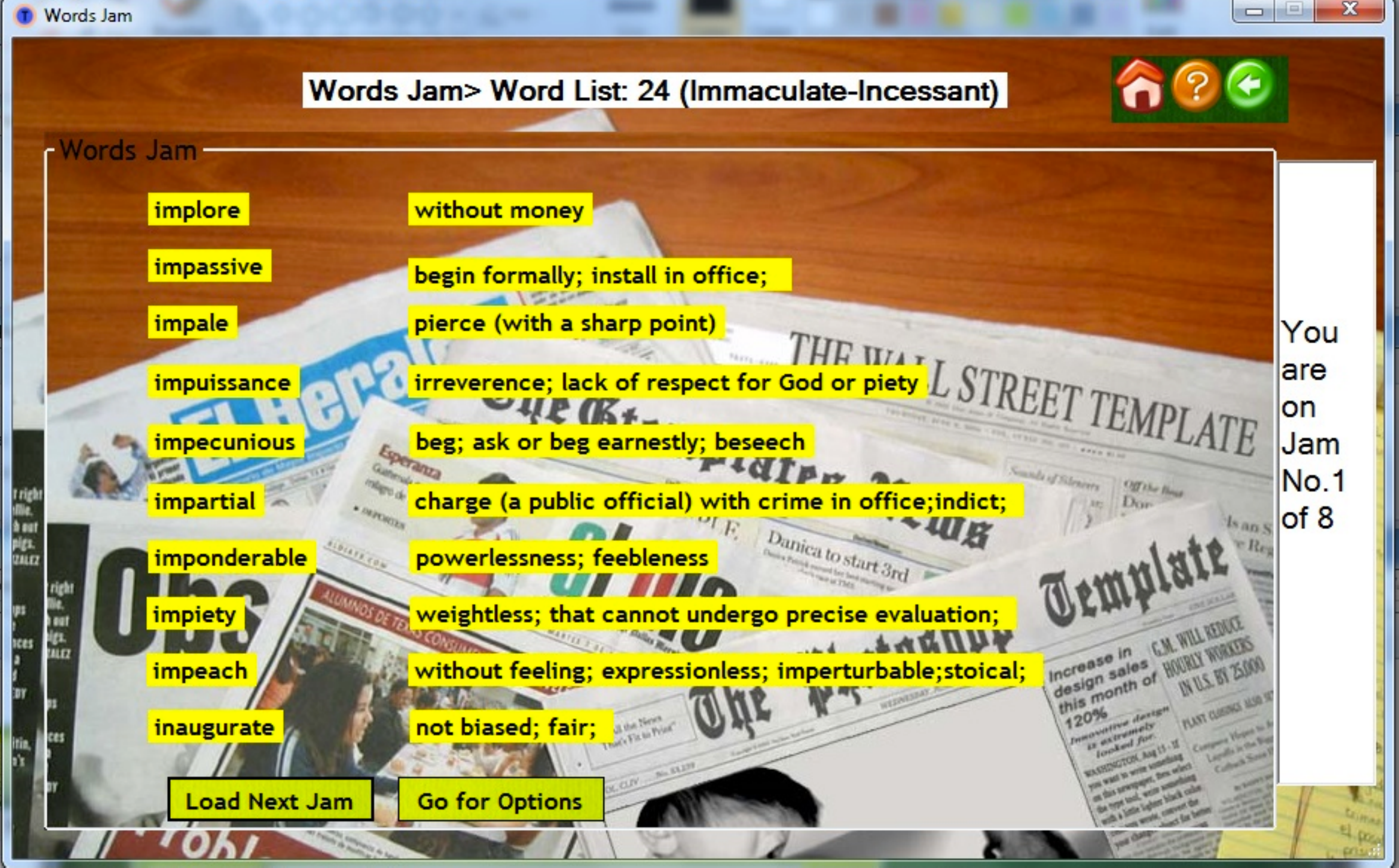}
\caption{Words jam feature in Tarantula.}
\label{fig:jam}
\end{figure}

\begin{figure}[htbp]
\centering
\includegraphics[height=5.50cm]{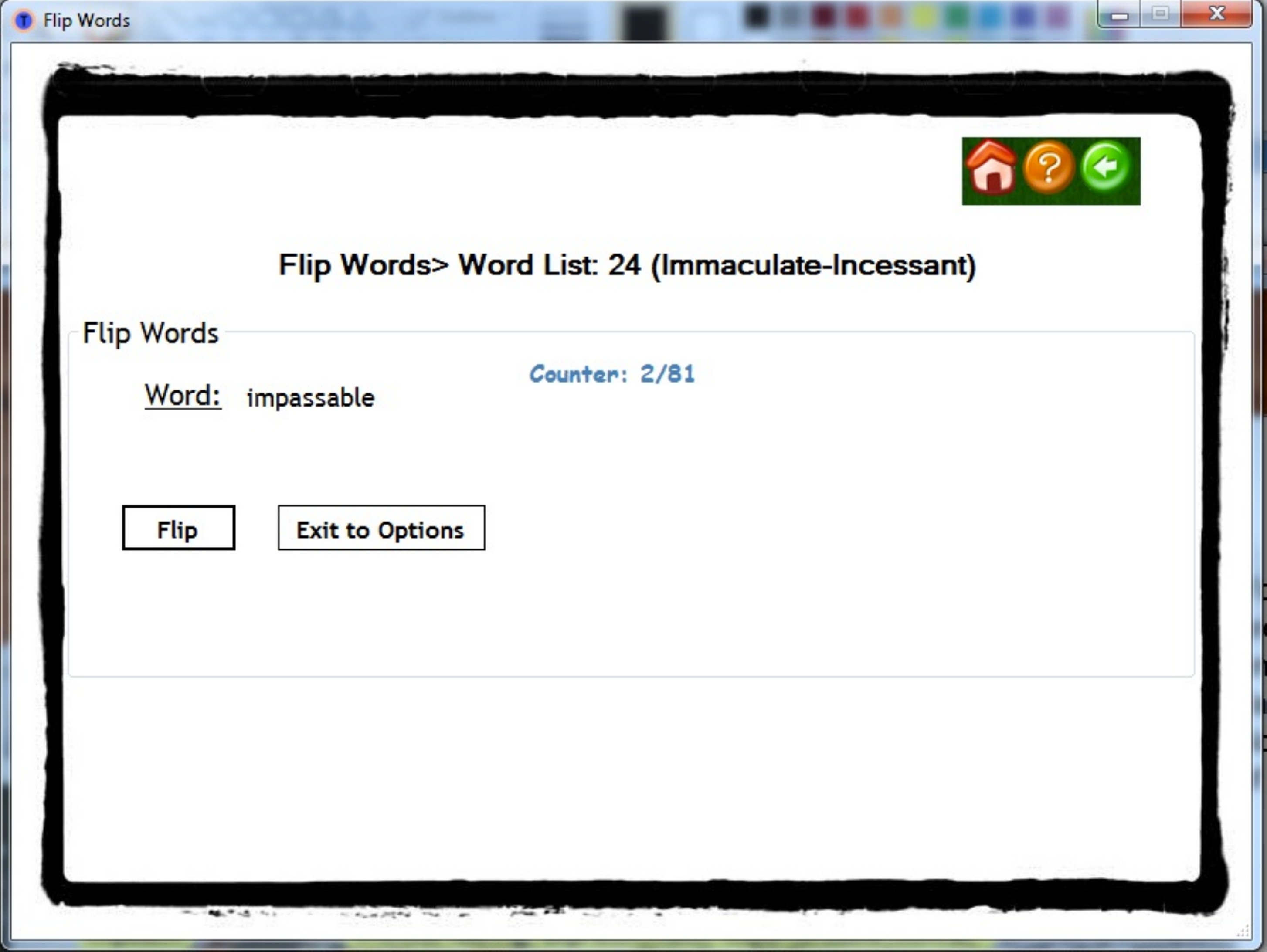}
\caption{Flip Words feature in Tarantula.}
\label{fig:flip}
\end{figure}

The ``Words Jam'' feature shows ten words from the word list on one side and their meanings from the word list on the other side.
Words' and their meanings' positions on each side is random (see Figure~\ref{fig:jam}).
Users have to click a word and then the meaning of the word (or vice versa).
If the meaning of the word is right then both of them will be disappeared from the Jam.
Users can load a new Jam by clicking ``Load Next Jam'' button.
A counter shows the number of the Jam a user is practicing.
 The ``Flip Words" feature shows a random word from the word list to guess (see Figure~\ref{fig:flip}).
 Users can click ``Flip'' button to see the meaning of the word.
 Users can use ``Next button'' for a new word.
 A counter also counts the number of words the user has seen.

\subsection{Development of Tarantula}
The Tarantula is a desktop application, written in C\# programming language and we used Microsoft Visual Studio $2010$ platform.
The software consists of $19390$ lines of code.
It can be installed from GitHub~\footnote{https://github.com/ImrulKayes/Tarantula/\\blob/master/Tarantula1.0.msi}.
The code of Tarantula is also publicly available at GitHub~\footnote{https://github.com/ImrulKayes/Tarantula}.

We first developed a web crawler using Python programming language to crawl  a subset of  HTML pages from~\cite{WordHacker2006list}.
These pages have all the words and their meanings.
We parsed the crawled HTML pages using a Parser (also written in Python).
The Parser went through all HTML pages and extracted words and meanings using regular expressions.
Then we created the repository of fifty word lists ($4054$ words) from the extracted words and meanings.
Finally, we used the repository as a word database for Tarantula.

\subsection{Test Cases and Faults}
\label{testcaseandfaults}
Based on the required features, we wrote fifty test cases before the development.
We ran the test cases after finishing an iteration of the development cycle.
Sixteen of the test cases revealed twenty three faults.
The test cases and the faults are below.

\begin{itemize}
\item {\textbf{Test Case \#$1$}\\
Action: click on a Word List icon to enable the system to load the words of the list with their meanings.\\
Expected result: the Word List should be loaded with features.\\
Fault \#$1$: the Word List is unavailable due to missing of the file.\\}

\item {\textbf{Test Case \#2}\\
Action: select the \emph{Learn Word List} option from the Radiobuttons of a word list.\\
Expected result: a random word from the word list and its meaning should be shown.\\
Fault \#$2$: the word in the selected word list is not generated.\\
Fault \#$3$: the meaning in the selected world list is not available.\\} 

\item {\textbf{Test Case \#$3$}:\\ 
Action: click on the  \emph{Next} button on the \emph{Learn Word List} feature.\\
Expected result: a new random word from the word list and its meaning should be shown.\\
Fault \#$4$: \emph{Next} button  does not generate a random word.\\
Fault \#$5$: \emph{Next} button does not generate a meaning.\\}

\item {\textbf{Test Case \#$4$}: \\
Action: Click on the \emph{Previous} button event in the \emph{Learn Word List} feature.\\
Expected result: a new random word from the word list and its meaning should be shown.\\
Fault \#$6$: \emph{Previous} button does not generate a random word.\\
Fault \#$7$: \emph{Previous} button does not generate a meaning.\\}

\item {\textbf{Test Case \#$5$}: \\
Action: check the \emph{Count} functionality in the  \emph{Learn Word List} feature by clicking on the \emph{Next} and \emph{Previous} buttons.\\
Expected result: \emph{Count} should be increased by one on clicking \emph{Next}  button and count should be decreased by one on clicking \emph{Previous}  button.\\
Fault \#$8$: \emph{Count} does not increase after clicking the \emph{Next} button.\\
Fault \#$9$: \emph{Count} does not decrease after clicking the \emph{Previous} button.\\}

\item {\textbf{Test Case \#$6$}: \\
Action:  select the \emph{Multiple Choice} option from the Radiobuttons of a word list.\\
Expected result: a random word from the wordlist and its possible choices of meanings should be shown.
The meanings are also from the same wordlist.\\
Fault \#$10$: the word  is not generated.\\
Fault \#$11$: meanings are not available.\\}

\item {\textbf{Test Case \#$7$}: \\
Action: verify the functionality of the \emph{Multiple Choice} option. 
Select the right meaning of the word. 
Select a wrong meaning of the word.\\
Expected result: the system should show ``Correct'' if the choice is right, otherwise it will show a message saying that the choice is wrong.\\ 
Fault \#$12$: the ``wrong'' message is not shown.\\}

\item {\textbf{Test Case \#$8$}: \\
Action: verify  \emph{Timer} functionality of the \emph{Multiple Choice} option.
Select the \emph{Multiple Choice} option from the Radiobuttons of a word list.
Then click the \emph{Next} button.\\
Expected result: The \emph{Timer} should start from a zero value.
It will increase by one after each second.
Clicking the \emph{Next} button should set it  a zero value.\\
Fault \#$13$: the \emph{Counter} does not increase.\\}

\item {\textbf{Test Case \#$9$}: \\
Action: select the \emph{Words Jam} option from the Radiobuttons of a word list.\\
Expected result: ten words and their meaning should be shown for matching from the word list.\\
Fault \#$14$: words in \emph{Words Jam}  are missing.\\
Fault \#$15$: meanings in \emph{Words Jam}  are missing.\\}

\item {\textbf{Test Case \#$10$}: \\
Action: click a word and then click its meaning in the \emph{Words Jam} feature.
Click a meaning and then click it's corresponding word in the \emph{Words Jam} feature.\\
Expected result: the word and the meaning should be disappeared.\\
Fault \#$16$: the word does not disappear.\\}

\item {\textbf{Test Case \#$11$}: \\ 
Action: in \emph{Words Jam} feature, click a word and click a wrong meaning of the word.\\
Expected result: the word and the meaning  should not be disappeared.\\
Fault \#$17$: the word disappears.\\}

\item {\textbf{Test Case \#$12$}: \\ 
Action: click \emph{Load Next Jam} in \emph{Words Jam} feature.\\
Expected result: ten words and their meanings should be shown to match and \emph{Jam counts} should be increased by one.\\
Fault \#$18$: \emph{Jam Count} does not increase.\\}

\item {\textbf{Test Case \#$13$}: \\ 
Action: select \emph{Flip Words} option from the Radiobuttons of a word list.\\
Expected result: a random word should be shown which will allow the users to guess the meaning of the word.\\
Fault \#$19$: the word is not generated.\\}

\item {\textbf{Test Case \#$14$}: \\  
Action: click the \emph{Flip} button in \emph{Flip Words} feature.\\
Expected result: The meaning of the word should be shown and the text ``Flip'' of the button should be changed as ``Next''.\\
Fault \#$20$: meaning is not available.\\
Fault \#$21$: text does not change.\\}

\item  {\textbf{Test Case \#$15$}: \\ 
Action: check the Radom word list generator, click the \emph{Rand} button.\\
Expected result: a random word list number should be generated.\\
Fault \#$22$: the random generator does not generate a random number.}\\

\item  {\textbf{Test Case \#$16$}: \\ 
Action: click the \emph{Go to Word List} button of the random wordlist generator.\\
Expected result:  the word list should be loaded with features.\\
Fault \#$23$: the word list is not loaded.}\\
\end{itemize}

\subsection{Properties of the Fault Network}
As discussed in Section~\ref{comreg}, a fault dependency network is a directed graph $F=(V,E)$, where a node $v \in V$ is a fault and an edge $e_{ij} \in E$ from $v_i \in V$ to $v_j \in V$ denotes that the fault $v_i$ is dependent on the fault $v_j$.
The directed graph can be represented by a $n * n$ matrix $F_{n*n}$, where an entry $F(i,j):$

\begin{equation}
F(i,j)=\begin{cases}
   1  &\text{if $e_{ij} \in E$}\\
    0, & \text{otherwise}
  \end{cases}
\end{equation}

\begin{figure}[htbp]
\centering
\includegraphics[height=8.00cm]{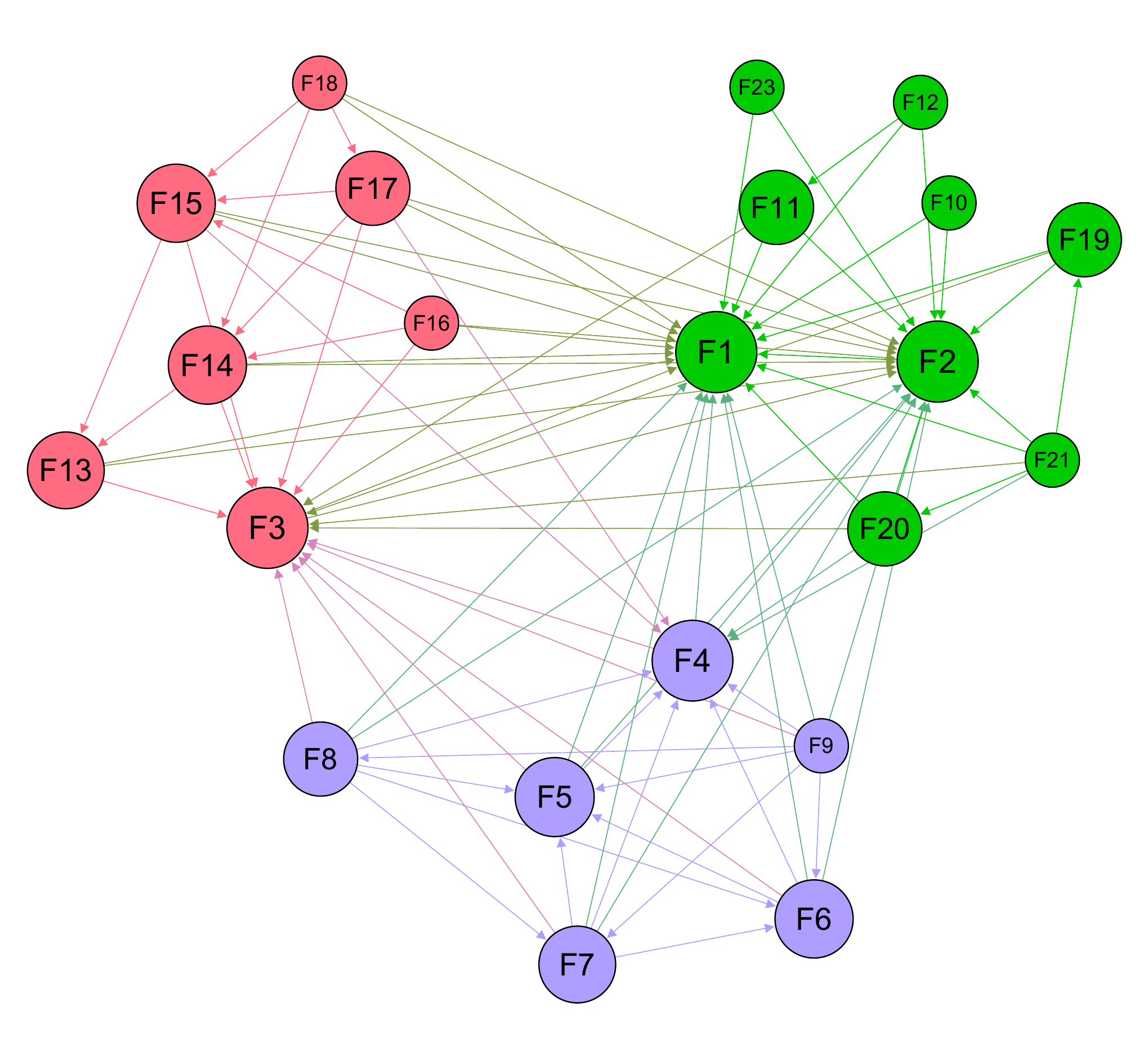}
\caption{The fault dependency network. Node size is proportional to in-degree. Nodes in a community are colored the same.}
\label{fig:fdg}
\end{figure}

The fault dependency matrix can be constructed after the system testing is done.
For example, in a Scrum process, a fault review is usually done before the regression testing by examining  reported  faults on the Dashboard.
In our case study, running the test cases we have got a fault dependency matrix $F$ shown in Table~\ref{tab:faultmatrix}.
The fault dependency matrix has $23$ faults and we associated relevant dependencies from~\ref{testcaseandfaults}.
Figure~\ref{fig:fdg} shows the largest component of the fault network ($22$ faults), where node size is proportional to in-degree of the node.
We used Gephi (\texttt{https://gephi.org/}) to visualize and obtain  structural properties of the network.
The structural properties of the fault network (largest component) are presented in Table~\ref{tab:summary}.

\begin{sidewaystable}
    \centering
\begin{tabular}{|c|c|c|c|c|c|c|c|c|c|c|c|c|c|c|c|c|c|c|c|c|c|c|c|c|c|}
\hline
  Faults$\downarrow$$\rightarrow$ &1 & 2& 3 &4 & 5& 6 &7 & 8 & 9 & 10 & 11& 12 & 13 & 14& 15 & 16 & 17& 18 & 19& 20& 21 & 22& 23\\
 \hline
1 &0 & 0& 0 &0 & 0& 0 &0 & 0& 0 &0 & 0& 0 &0 & 0& 0 &0 & 0& 0 & 0& 0& 0 & 0 &0\\
2 &1 & 0& 0 &0 & 0& 0 &0 & 0& 0 &0 & 0& 0 &0 & 0& 0 &0 & 0& 0 & 0& 0& 0 & 0 &0\\
3 &1 & 1& 0 &0 & 0& 0 &0 & 0& 0 &0 & 0& 0 &0 & 0& 0 &0 & 0& 0 & 0& 0& 0 & 0 &0\\
4 &1 & 1& 1 &0 & 0& 0 &0 & 0& 0 &0 & 0& 0 &0 & 0& 0 &0 & 0& 0 & 0& 0& 0 & 0 &0\\
5 &1 & 1& 1 &1 & 0& 0 &0 & 0& 0 &0 & 0& 0 &0 & 0& 0 &0 & 0& 0 & 0& 0& 0 & 0 &0\\
6 &1 & 1& 1 &1 & 1& 0 &0 & 0& 0 &0 & 0& 0 &0 & 0& 0 &0 & 0& 0 & 0& 0& 0 & 0 &0\\
7 &1 & 1& 1 &1 & 1& 1 &0 & 0& 0 &0 & 0& 0 &0 & 0& 0 &0 & 0& 0 & 0& 0& 0 & 0 &0\\
8 &1 & 1& 1 &1 & 1& 1 &1 & 0& 0 &0 & 0& 0 &0 & 0& 0 &0 & 0& 0 & 0& 0& 0 & 0 &0\\
9 &1 & 1& 1 &1 & 1& 1 &1 & 1& 0 &0 & 0& 0 &0 & 0& 0 &0 & 0& 0 & 0& 0& 0 & 0 &0\\
10 &1 & 1& 0 &0 & 0& 0 &0 & 0& 0 &0 & 0& 0 &0 & 0& 0 &0 & 0& 0 & 0& 0& 0 & 0 &0\\
11 &1 & 1& 1 &0 & 0& 0 &0 & 0& 0 &0 & 0& 0 &0 & 0& 0 &0 & 0& 0 & 0& 0& 0 & 0 &0\\
12 &1 & 1& 0 &0 & 0& 0 &0 & 0& 0 &0 & 1& 0 &0 & 0& 0 &0 & 0& 0 & 0& 0& 0 & 0 &0\\
13 &1 & 1& 1 &0 & 0& 0 &0 & 0& 0 &0 & 0& 0 &0 & 0& 0 &0 & 0& 0 & 0& 0& 0 & 0 &0\\
14 &1 & 1& 1 &0 & 0& 0 &0 & 0& 0 &0 & 0& 0 &1 & 0& 0 &0 & 0& 0 & 0& 0& 0 & 0 &0\\
15 &1 & 1& 1 &1 & 0& 0 &0 & 0& 0 &0 & 0& 0 &1 & 0& 0 &0 & 0& 0 & 0& 0& 0 & 0 &0\\
16 &1 & 1& 1 &0 & 0& 0 &0 & 0& 0 &0 & 0& 0 &0 & 1& 1 &0 & 0& 0 & 0& 0& 0 & 0 &0\\
17 &1 & 1& 1 &1 & 0& 0 &0 & 0& 0 &0 & 0& 0 &0 & 1& 1 &0 & 0& 0 & 0& 0& 0 & 0 &0\\
18 &1 & 1& 0 &0 & 0& 0 &0 & 0& 0 &0 & 0& 0 &0 & 1& 1 &0 & 1& 0 & 0& 0& 0 & 0 &0\\
19 &1 & 1& 1 &0 & 0& 0 &0 & 0& 0 &0 & 0& 0 &0 & 0& 0 &0 & 0& 0 & 0& 0& 0 & 0 &0\\
20 &1 & 1& 1 & 1& 0 &0 & 0& 0 &0 & 0& 0 &0 & 0& 0 &0 & 0& 0 & 0& 0& 0 & 0 &0&0\\
21 &1 & 1& 1 & 1& 0 &0 & 0& 0 &0 & 0& 0 &0 & 0& 0 &0 & 0& 0 & 0& 1& 1 & 0 &0&0\\
22 &0 & 0& 0 &0 & 0& 0 &0 & 0& 0 &0 & 0& 0 &0 & 0& 0 &0 & 0& 0 & 0& 0& 0 & 0 &0\\
23 &1 & 1& 0 &0 & 0& 0 &0 & 0& 0 &0 & 0& 0 &0 & 0& 0 &0 & 0& 0 & 0& 0& 0 & 0 &0\\
\hline
\end{tabular}
\captionof{table}{Fault dependency matrix.}
\label{tab:faultmatrix}
\end{sidewaystable}

\begin{table}[htbp]
\centering
\begin{tabular}{|l|r|}
\hline
\textbf{Nodes}&$22$\\
\hline
\textbf{Edges}&$97$\\
\hline
\textbf{Average In-degree}&$3.95$\\
\hline
\textbf{Average Path length}&$1.074$\\
\hline
\textbf{Average Clustering Coefficient}&$0.416$\\
\hline
\end{tabular}
\captionof{table}{Structural properties of the fault network.}
\label{tab:summary}
\end{table}

A notable characteristic of the fault network is the high clustering coefficient. 
Given a network $G = (V, E)$, the clustering coefficient $C_i$ of a node $i \in V$ is the proportion of all the possible edges between neighbors of the node that actually exist in the network~\cite{Newman2010Book}. 
The clustering coefficient is defined as:

\begin{equation}
C=\frac{3* \text{Number of triangles}}{\text{Number of connected triples of the nodes}}
\end{equation}

In the fault network, for a node $v_i \in V$ there could be $k_i(k_i-1)$  links exist among  the neighborhood of $v_i$, where $k_i$ is the number of neighbor of $v_i$.
So, the local clustering coefficient of fault $v_i$ in the fault network is:
\begin{equation}
C=\frac{| \{e_{jk}:v_j,v_k  \in \text{Neighbor}(V_i), e_{jk} \in E\}|}{k_i*(k_i-1)}
\end{equation}

The clustering coefficient for the whole network is  the average of the local clustering coefficients of all the nodes $n$~\cite{Watts1998Nature}:

\begin{equation}
C = \frac{1}{n}\sum_{i=1}^{n} C_i
\end{equation}

A high clustering coefficient in the fault networks implies that a faults's connections are interconnected and have a greater effect on one another. 
The small average path length ($1.074$), comparable with that of the corresponding random graph of the same size ($1.675$), together with the high average clustering coefficient (the fault network has average clustering coefficient $0.416$, where a same size random graph has $0.295$), places the fault network in the category of small-world graphs~\cite{Watts1998Nature}.

\subsection{Leading Faults and Prioritized Test Cases}
\begin{figure*}[ht]
\begin{minipage}[htb]{0.50\linewidth}
\centering
\includegraphics[height=4.5cm]{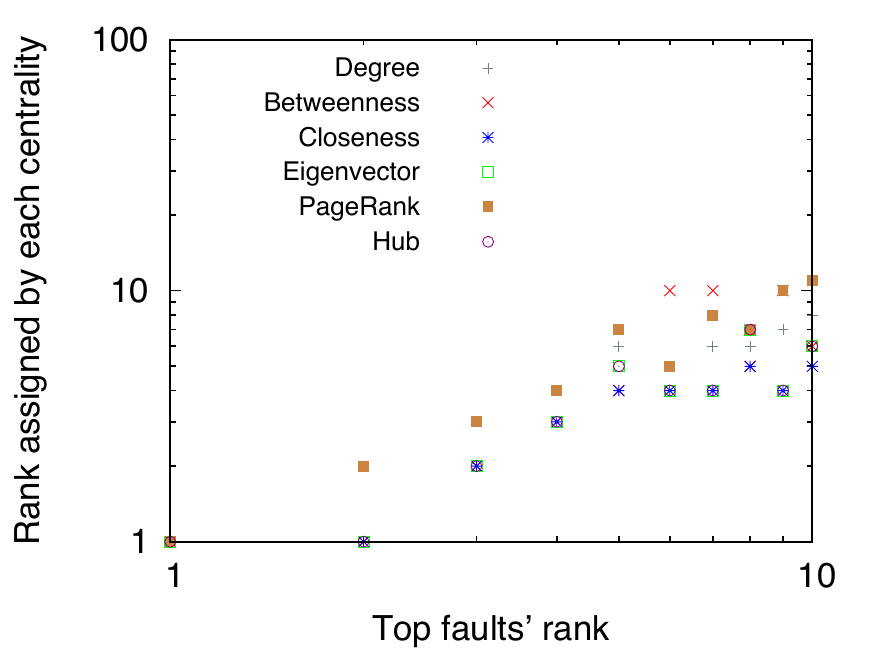}
\caption{Assigned rank of top ten most central faults from all centralities.}
\label{fig:overlap}
\end{minipage}
\hspace{.5cm}
\begin{minipage}[htb]{0.45\linewidth}
\captionof{table}{Average rank of the top ten faults.}
\centering
\begin{tabular}{c|c}
\hline
\textbf{Faults' ID} & \textbf{Average Rank} \\ 
\hline
$Fault \#1$	&$1.00$\\ 
$Fault \#2$ 	&$1.33$\\ 
$Fault \#3$ 	&$2.33$\\
$Fault \#4$ 	&$3.33$\\
$Fault \#15$	&$5.16$\\
$Fault \#5$	&$5.33$\\
$Fault \#6$	&$6.00$\\
$Fault \#14$	& $6.17$\\
$Fault \#7$	&$6.5$\\
$Fault \#17$	&$7.00$\\
\hline
\end{tabular}
\label{tab:topten}
\end{minipage}
\end{figure*}

As described in Section~\ref{method}, we use centrality metrics to rank leading faults in the network.
To manage the fault network and compute centralities, we used Python 2.7 with the NetworkX\footnote{http://networkx.github.io/} library. 
Faults that appear among the top $10$ in multiple centrality metrics are represented in color in Table~\ref{tab:comparison}.
The average rank of the top $10$ leading faults and their average ranks considering all centralities are shown in Table~\ref{tab:topten}.
Out of the $10$ leading faults listed on each centrality, $6$ faults (60\%) are common in all the centralities.  
To observe more closely, we plot the ranks of  top $10$ of the faults assigned by all centralities, showed in Figure~\ref{fig:overlap}.
As expected, a fault's assigned ranks from centralities form a cluster and together with all the clusters we can visualize a straight line.
This shows that all the centralities tend to rank the same fault in the top.

Note Pareto principle described in Section~\ref{method}---for many events, roughly 80\% of the effects come from 20\% of the causes.
Pareto principle also holds for the fault dependency network.
In the fault dependency network, $78$ out of $97$ ($80.41\%$) edges are incident on top $5$ nodes out of $23$ nodes (21.73\%).
It shows that $80.41\%$ of the fault dependencies are due to 21.73\% of faults---almost equal figures from Pareto principle.

\begin{sidewaystable}
    \centering
\captionof{table}{Top $10$ faults according to each centrality measurement, sorted in increasing order by rank from left to right. IDC: in-degree centrality, BC: betweenness centrality, CC: closeness centrality, EC: eigenvector centrality, PG: page-rank centrality, and HC: hub centrality. Faults common to all centralities are colored the same.}
\begin{tabular}{|c|c|c|c|c|c|c|c|c|c|c|}
\hline

\textbf{IDC}&\textcolor{red}{$Fault \#1$}&\textcolor{blue}{$Fault \#2$}&\textcolor{violet}{$Fault \#3$}&\textcolor{orange}{$Fault \#4$}&\textcolor{cyan}{$Fault \#5$}& \textcolor{brown}{$Fault \#14$}& \textcolor{magenta}{$Fault \#15$} &\textcolor{teal}{$Fault \#6$}& \textcolor{olive}{$Fault \#13$}&\textcolor{green}{$Fault \#7$}\\

\textbf{BC}&\textcolor{red}{$Fault \#1$}&\textcolor{blue}{$Fault \#2$}&\textcolor{violet}{$Fault \#3$}&\textcolor{orange}{$Fault \#4$}& \textcolor{magenta}{$Fault \#15$} &\textcolor{brown}{$Fault \#14$}&\textcolor{green}{$Fault \#17$}&$Fault \#21$&$Fault \#21$&\textcolor{olive}{$Fault \#13$}\\


\textbf{CC}&\textcolor{red}{$Fault \#1$}&\textcolor{blue}{$Fault \#2$}&\textcolor{violet}{$Fault \#3$}&\textcolor{orange}{$Fault \#4$}&\textcolor{magenta}{$Fault \#15$}&\textcolor{cyan}{$Fault \#5$}&\textcolor{teal}{$Fault \#6$}&\textcolor{green}{$Fault \#7$}&\textcolor{olive}{$Fault \#8$}&\textcolor{blue}{$Fault \#9$}\\

\textbf{EC}&\textcolor{red}{$Fault \#1$}&\textcolor{blue}{$Fault \#2$}&\textcolor{violet}{$Fault \#3$}&\textcolor{orange}{$Fault \#4$}&\textcolor{cyan}{$Fault \#5$}&\textcolor{teal}{$Fault \#6$}&\textcolor{green}{$Fault \#7$}&\textcolor{olive}{$Fault \#8$}&\textcolor{blue}{$Fault \#9$}&\textcolor{magenta}{$Fault \#15$}\\

\textbf{PC}&\textcolor{red}{$Fault \#1$}&\textcolor{blue}{$Fault \#2$}&\textcolor{violet}{$Fault \#3$}&\textcolor{orange}{$Fault \#4$}&\textcolor{cyan}{$Fault \#5$}& \textcolor{olive}{$Fault \#13$}&\textcolor{brown}{$Fault \#14$}&\textcolor{magenta}{$Fault \#15$}&\textcolor{teal}{$Fault \#6$}& $Fault \#11$\\

\textbf{HC}&\textcolor{red}{$Fault \#1$}&\textcolor{blue}{$Fault \#2$}&\textcolor{violet}{$Fault \#3$}&\textcolor{orange}{$Fault \#4$}&\textcolor{cyan}{$Fault \#5$}&\textcolor{teal}{$Fault \#6$}&\textcolor{green}{$Fault \#7$}&\textcolor{olive}{$Fault \#8$}&\textcolor{blue}{$Fault \#9$}& \textcolor{magenta}{$Fault \#15$}\\
\hline
\end{tabular}
\label{tab:comparison}
\end{sidewaystable}

Finally, our prioritized ordering of the test cases for regression testing based on leadings faults' exposure in test cases is:
T1, T2, T3, T4, T9, T11, T5, T8, T14, T6, T10, T12, T13, T7, T16, T15 (T denotes a test case).

\subsection{Effectiveness of ComReg}

We used three techniques to prioritize our regression test suite and compared them to ComReg.
We want to observe which method has faster  fault dependency detection rate.
The techniques are the following:

\begin{enumerate}
  \item Prioritization using relevant slices (ReSl): ReSl prioritizes test cases  taking into account the coverage requirements present in the relevant slices of the outputs of test cases~\cite{Jeffrey2006Slice}.
  \item Prioritization  based on Function Call Path (FuCa):  FuCa leverages function call-level paths and prioritizes test cases based on those coverage paths~\cite{Zhang2012Function}.
  \item Random Prioritization (Random): Random prioritizes test cases based  on a randomization algorithm. 
\end{enumerate}

\begin{figure}[htbp]
\centering
\includegraphics[height=6.00cm]{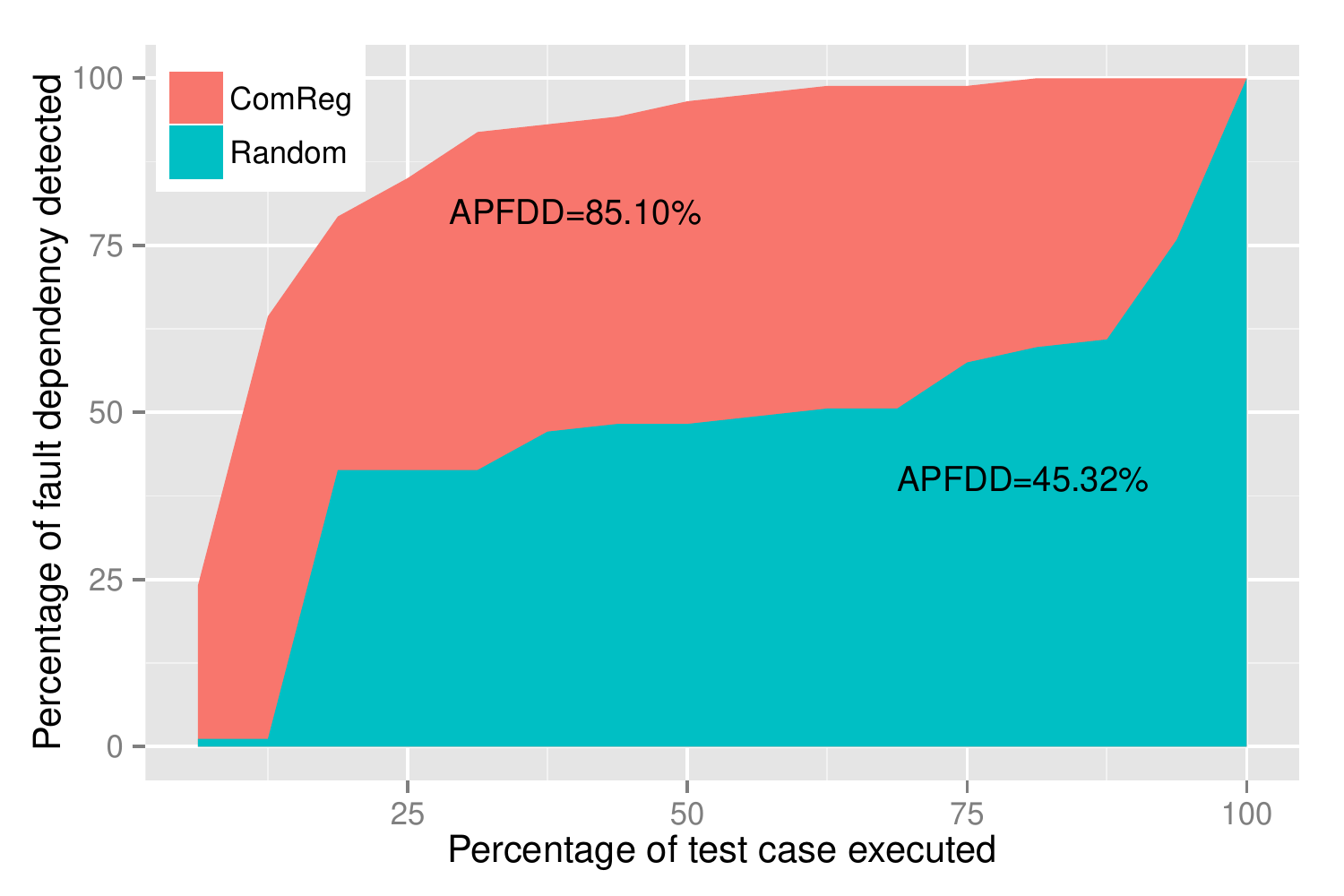}
\caption{Average percentage of fault dependency detected (APFDD) for the prioritized test cases using ComReg and random techniques.}
\label{fig:network_random}
\end{figure}

\begin{figure}[htbp]
\centering
\includegraphics[height=6.00cm]{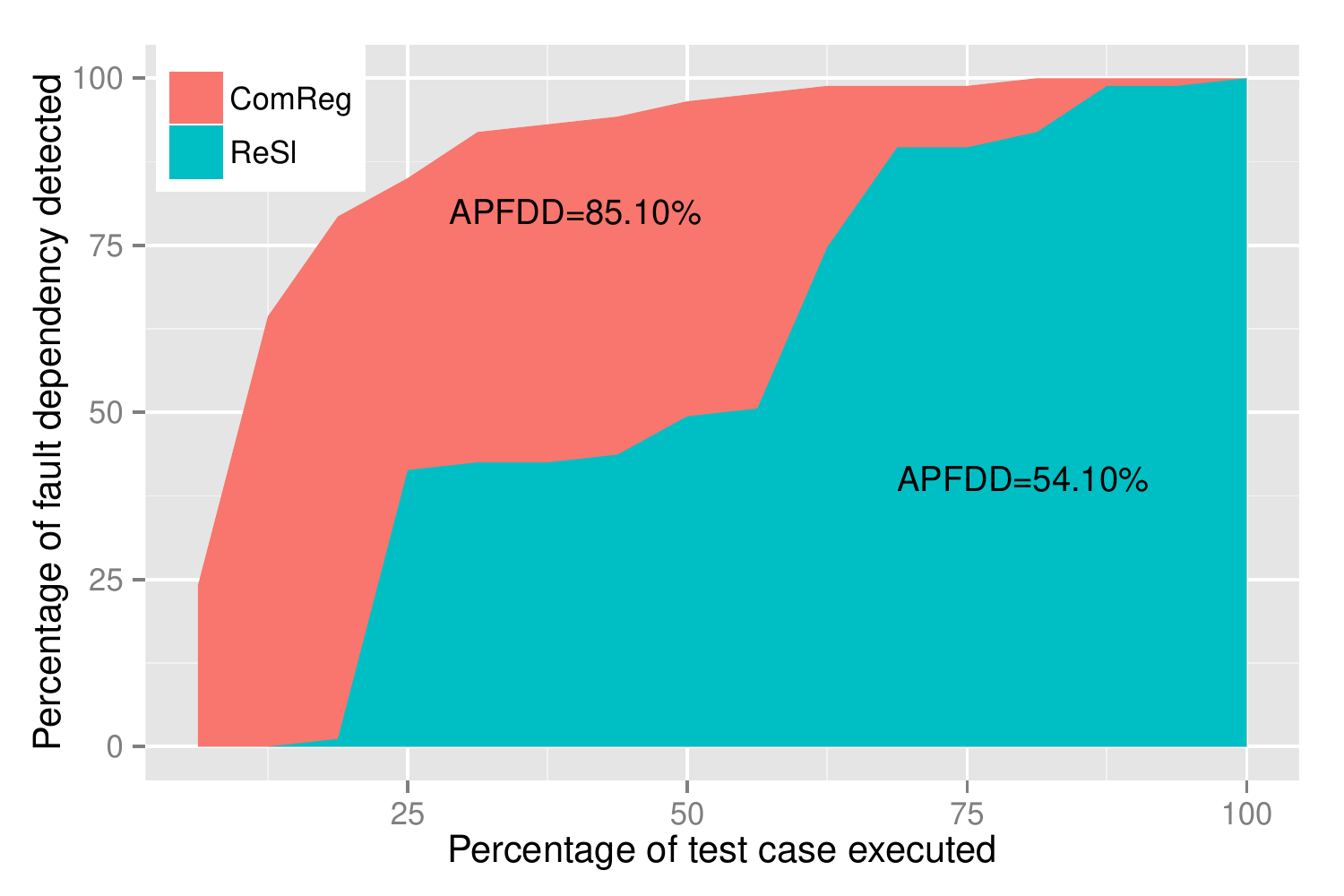}
\caption{Average percentage of fault dependency detected (APFDD) for the prioritized test cases using ComReg and ReSl techniques~\cite{Jeffrey2006Slice}.}
\label{fig:network_ReSl}
\end{figure}

\begin{figure}[htbp]
\centering
\includegraphics[height=6.00cm]{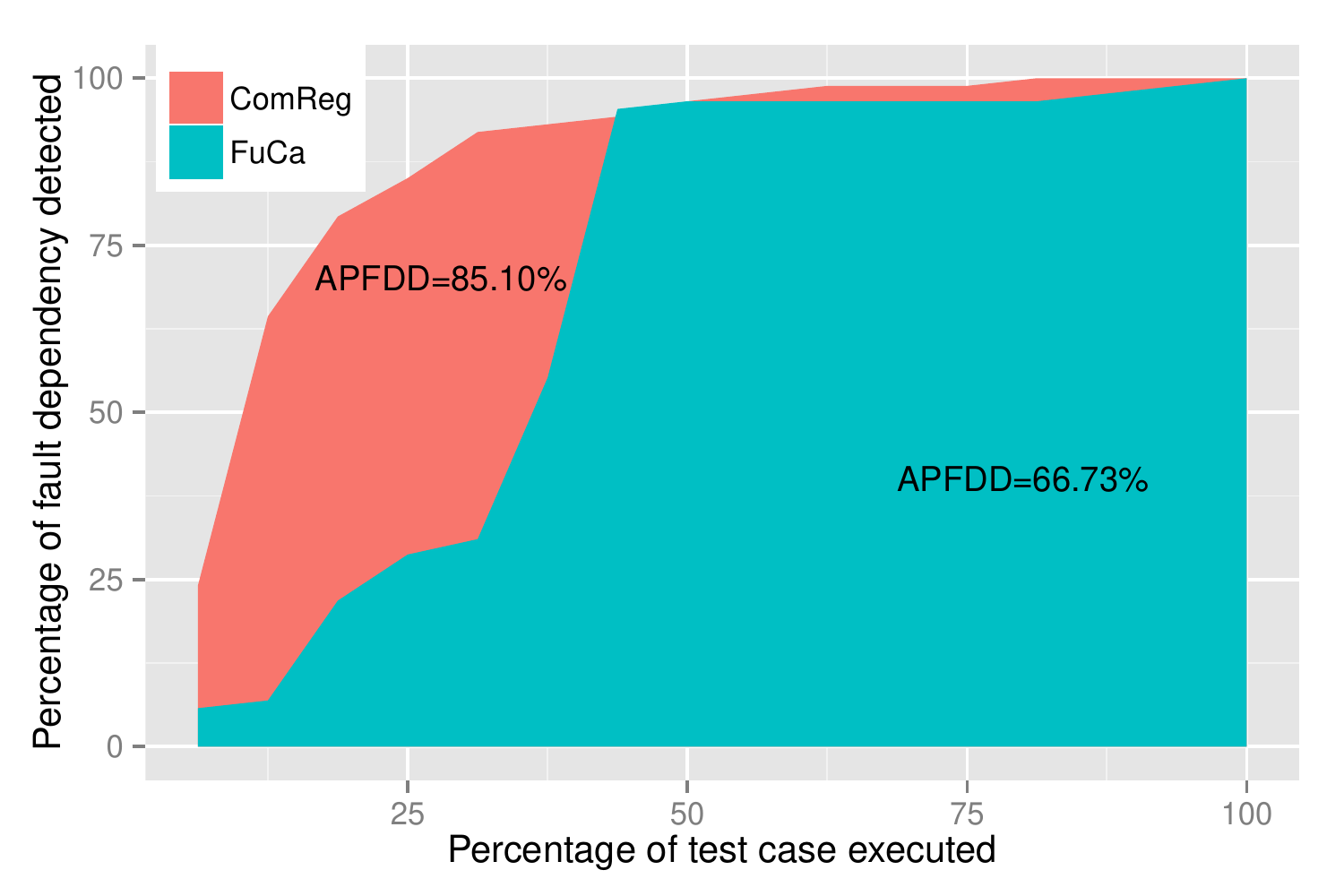}
\caption{Average percentage of fault dependency detected (APFDD) for the prioritized test cases using ComReg and Function Call Path techniques~\cite{Zhang2012Function}.}
\label{fig:network_coverage}
\end{figure}

We used  a metric, APFDD (Average of the Percentage Fault Dependency Detected)~\cite{Kayes2011Fault}, to measure effectiveness of ComReg to the techniques described above.
The APFDD quantifies how rapidly a prioritized test suite can detect dependency among faults
The values of the APFDD range from 0 to 100; higher value implies faster fault dependency detection.
Figures~\ref{fig:network_random},~\ref{fig:network_ReSl},~\ref{fig:network_coverage} show the percentage of test cases executed and the percentage of fault dependency detected for the test cases prioritized by ComReg and other methods (random, ReSl and FuCa respectively).
The areas under the curves represent the weighted average of the percentage of the fault dependency detected (APFDD).
From the figures we see that the random prioritization method performed the worst, yielded only 45.32\% APFDD.
The ReSl and FuCa methods performed moderately, both were better than the  random prioritization method with APFDD 54.10\% and 66.73\% respectively.
Our method ComReg provided the best value of the APFDD (85.10\%), hence outperformed the random, ReSl and FuCa methods in rapidly detecting fault dependencies.

\subsection{Community Detection Techniques}

 We used a popular modularity maximization approach, Louvain method~\cite{blondel2008fast}, to detect fault communities in the network.
Louvain method is a greedy optimization method that attempts to optimize the modularity of a partition of the fault network.
The optimization is performed in two steps: modularity maximization and community aggregation. 
In modularity maximization step, the method looks for  ``small'' communities by optimizing modularity locally. 
In community aggregation step, it aggregates nodes who belong to the same community and builds a new network whose nodes are the communities. 
Two steps are repeated iteratively until a maximum of modularity is attained and a hierarchy of communities is produced.
Applying the algorithm on the fault network, we have got three fault communities: community \#1 (pink color nodes in Figure~\ref{fig:fdg}) has seven faults, community \#2 (green color nodes in Figure~\ref{fig:fdg}) has nine faults and community \#3 (violet color nodes in Figure~\ref{fig:fdg}) has six faults.
Leading faults are distributed among communities.
For example, community \#1, community \#2 and community \#3 have one, two and two top leading faults respectively out of top five leading faults.
As discussed in the Section~\ref{method}, leading faults and faults from their communities revealed by the test cases (which target prioritized features) could be used in selecting regression test cases.
For example, leading fault Fault \# 1's community has eight faults originated from seven test cases (46.66\% of test cases).


\section{Related Work}
\label{related}

Different solutions have been proposed to prioritize test cases for regression testing. 
In this section, we discuss test case prioritization techniques from the literature.

\emph{Coverage-based prioritization} techniques aim to achieve higher fault detection rates by maximizing early coverage.
The solutions are inspired by the intuition that early maximization of structural coverage will also maximize early fault detection.
Rothermel et al. proposed a family of techniques~\cite{Rothermel1999Prio,Rothermel2001Pri} for test-case prioritization based on several coverage criteria.
They considered different types of coverages: branch-total, branch-additional, statement-total, statement-additional, Fault Exposing Potential (FEP)-total, and FEP-additional.
A branch-total coverage solution prioritizes test cases according to the number of branches covered by individual test cases.
On the other hand, branch-additional prioritizes test cases according to the additional number of branches covered by individual test cases.
Statement-total and statement-additional coverage based solutions are similar to previous two approaches, but rather than considering branches, they consider statements.
The FEP-total and  FEP-additional are based on program mutation.
Program mutation produce a mutant version of the program by introducing modifications to the program source. 
The prioritization techniques prioritize the test cases such that the test cases can reveal the difference between the original program and the mutant. 
The authors introduced a metic Average Percentage of Fault Detection (APFD) to quantify the success of a prioritization.
Elbaum et al.~\cite{Elbaum2000PTC,Elbaum2002Pri} further proposed prioritization techniques covering coverage criterion at the function level, while Do et al.~\cite{Do2004Pri}  considered the coverage criteria at the block level.
Korel et al. discussed several model-based test prioritization heuristics in~\cite{Korel2005Test,Korel2008System}.
Their coverage criteria is system model; they identified elements of the model related to source-code modifications and applied heuristics to prioritize test cases so that early fault detection in the modified system is maximized.
Jones and Harrold described a fine-grain coverage criterion in ~\cite{Jones2003Fine}, which considers a modified condition/decision coverage.

\emph{Requirement-based approaches} consider a software's requirements as a basis for prioritization of test cases.
Srikanth et al.~\cite{Srikanth2005Requirement} prioritized test cases  based on four factors: requirements volatility, customer priority, implementation complexity, and fault proneness of the requirements.
Krishnamoorthi at al.~\cite{krishnamoorthi2009factor} adopted a similar approach.
Their prioritization is based on six factors: customer priority, changes in requirement, implementation complexity, completeness, traceability and fault impact. 
However, a potential weakness of requirement-based approaches is that requirement properties are subjective and thus estimations might be biased.

\emph{Constraint-based approaches} consider different constraints and practical complications in test case prioritization.
Kim et at.~\cite{Kim2002Cons} consider resource and time constraints.
The resource and time constraint do not allow the execution of the entire test suite for a regression testing.
They proposed a  heuristic that uses historical information to do test case  prioritization.
Alspaugh et al.~\cite{Alspaugh2007ETP}  consider a situation when regression testing is  performed in a time constrained environment.
They  empirically compared seven Knapsack solvers (e.g., greedy, dynamic programming and the core algorithm) and identified a test suite reordering that rapidly covers the test requirements and always terminates within a specified testing time limit.
Walcott et al.~\cite{Walcott2006TTS} proposed a genetic algorithm-based time-ware test case prioritization technique and empirically compared the  approach with the initial ordering, the reverse ordering and two control techniques (random prioritization and fault-aware prioritization). 
They  defined a metric to  evaluate the effectiveness of prioritization in a time-constrained environment.
Zhang et al.~\cite{Zhang2009TTP} also studied time-aware  test case  prioritization problem.
Their proposed test case prioritization is based on integer linear programming.
They empirically showed that their two proposed techniques outperform genetic algorithms-based time-aware test case prioritization and four other traditional techniques for test-case prioritization.

Researchers used a number of other criteria to prioritize test cases.
Sherriff et al. prioritized test cases based on historical change records in~\cite{Sherriff2007History}.
They proposed a methodology for determining the effect of a software feature change and then prioritized regression test cases by gathering software change records and analyzing them through singular value decomposition. 
Leon et al. ~\cite{Leon2003distribution} introduced distribution-based filtering  and prioritized test cases based on the distribution of the profiles of test cases in the multi-dimensional profile space. 
Sampath et al.~\cite{Sampath2008Web} prioritized test cases for web applications. 
They prioritized test suites by test lengths, frequency of appearance of request sequences and systematic coverage of parameter-values and their interactions.
Rummel et al.~\cite{Rummel2005Dataflow} introduced a prioritization technique based on data-flow analysis.
They focused on the definition and use of program variables for the data-flow analysis.
Jeffrey et al.~\cite{Jeffrey2006Slice} prioritized test cases using relevant slices.
Qu et al.~\cite{Qu2007Black} prioritized test cases in a black box environment.

However, none of the above solutions considered dependencies among faults in prioritizing test cases for regression testing.
In software testing, it is known that some faults  are the consequences of other faults (leading faults).
So, intuitively, test cases that revealed the leadings faults should be executed first in a regression testing in order to get an early confirmation that software is free from dependent faults.
In~\cite{Kayes2011Fault}  we took the first step to prioritize regression testing based on fault dependency.
We proposed an algorithm to prioritize test cases based on fault dependency.
We also proposed a metric Average Average Percentage Fault Dependency Detected (APFDD) to quantify how rapidly a prioritized test suite can detect dependencies among faults.
However, that work only considered $1$-hop neighborhood or dependencies of faults.
This paper leverages a fault network  for prioritization.

\section{Summary and Discussions}
\label{conclusion}

In this paper, we have presented ComReg, which uses a fault dependency network to prioritize test cases for regression testing. 
We have modeled a fault dependency network as a directed graph and identified leading faults to prioritize test cases. 
We have leveraged a network centrality aggregation technique in the fault dependency network to identify leading faults.
The centrality aggregation technique considers six representative centrality metrics such as indegree, betweenness, closeness, eigenvector, pagerank and hub centrality and offers a final leading score to identify the leading faults.
Our discussions on fault communities shed light on  selecting X\% of the test cases from a prioritized regression test suite.
Finally, we have presented a case study.
In the case study, we have developed an English vocabulary learning software, ``Tarantula'' and identified leading faults from a fault network after running a set of test cases at the end of the first phase of the development.
We have showed the fault communities in the fault network for test case selection from a prioritized regression test suite.

The fault dependency network might not be a connected graph.
For example, in our fault dependency network of Tarantula consists of two components.
However, small-world networks tend to have  giant components(e.g.,~\cite{Faloutsos1999Internet,Aiello2001Telephone,redner1998Citation}).
A giant component is a connected subgraph that contains a majority of the entire graph's nodes~\cite{newman2001graph}.
The giant component  fills most of the network---usually more than half and not infrequently over 90\%---while the rest of the network is divided into a large number of small components disconnected from the rest~\cite{Newman2010Book}.  
Our small-world fault dependency network also has one giant component ($22$ nodes).
So, if a fault network has a large number of nodes and if it shows a large number of connected components, the giant component could be leveraged to detect the leading faults.

Our work has multiple limitations.
First, we built a subject software (``Tarantula'') to present a case study and show the effectiveness of our prioritization technique.
The Tarantula is a medium-scale software, which lacks the rigorous development cycle of a typical commercial software.
This leads to a higher number of faults in system testing, even using  a small number of test cases.
Using our proposed method in an industrial software testing setting could provide more insights.

Second, we do not consider the time and resources (e.g., testers) required to identify fault dependencies.
If a software is poorly written with a lot of fault cascades, identifications of  fault dependencies and their management might be costlier than running the full test suite.

Finally, some  centrality algorithms (e.g., betweenness, closeness) used by ComReg are computationally expensive.
This was not a major issue for the small fault dependency graph discussed in this paper.
However, for a large-scale fault dependency graph, an approximation algorithm (e.g., k-path centrality~\cite{kpath}) with parallel implementation is required for efficiency.

\bibliographystyle{splncs}
\bibliography{Bibtex}

\end{document}